\documentclass[fleqn,usenatbib,useAMS]{mnras}

\usepackage{graphicx}	% Including figure files
\usepackage{amsmath}	% Advanced maths commands
\usepackage{amssymb}	% Extra maths symbols
\usepackage{multicol}        % Multi-column entries in tables
\usepackage{array}
\usepackage{multirow}
\usepackage{pdflscape}	% Landscape pages
\usepackage{xurl}
\usepackage{hyperref}
\usepackage{amsfonts}
\usepackage[utf8]{inputenc}
\usepackage{microtype}
\usepackage{siunitx}
\usepackage{soul}
\usepackage{longtable}
\usepackage{bm}
\usepackage{natbib}
\usepackage[normalem]{ulem}
\usepackage[table]{xcolor}
\usepackage{color}
\usepackage{comment}
\usepackage{widetext}
\usepackage[version=4]{mhchem}
\usepackage{xspace}
\usepackage{tabularray}
\DeclareRobustCommand{\VAN}[3]{#2}
\let\VANthebibliography\thebibliography
\def\thebibliography{\DeclareRobustCommand{\VAN}[3]{##3}\VANthebibliography}

\definecolor{forestgreen}{rgb}{0.13, 0.545, 0.13}

\providecommand{\tabularnewline}{\\}

\newcommand{\dunit}{\,cm$^{2}$\,s$^{-1}$\xspace}

\newcommand{\punit}{\,erg\,s$^{-1}$\xspace}

\newcommand{\gev}{\,GeV\xspace}
\newcommand{\tev}{\,TeV\xspace}
\newcommand{\pev}{\,PeV\xspace}

\def\refjnl#1{{\rm{#1}}}
\def\aj{\refjnl{AJ}}                   % Astronomical Journal
\def\araa{\refjnl{ARA\&A}}             % Annual Review of Astron and Astrophys
\def\apj{\refjnl{ApJ}}                 % Astrophysical Journal
\def\apjl{\refjnl{ApJ}}                % Astrophysical Journal, Letters
\def\apjs{\refjnl{ApJS}}               % Astrophysical Journal, Supplement
\def\ao{\refjnl{Appl.~Opt.}}           % Applied Optics
\def\aap{\refjnl{A\&A}}                % Astronomy and Astrophysics
\def\mnras{\refjnl{MNRAS}}             % Monthly Notices of the RAS
\def\prd{\refjnl{Phys.~Rev.~D}}        % Physical Review D
\def\prl{\refjnl{Phys.~Rev.~Lett.}}    % Physical Review Letters
\def\ssr{\refjnl{Space~Sci.~Rev.}}     % Space Science Reviews
\def\nat{\refjnl{Nature}}              % Nature
\usepackage{lineno}
\usepackage[T1]{fontenc}
\usepackage{ae,aecompl}
\graphicspath{{Figures/}}

\usepackage{newtxtext,newtxmath}
\title[Detecting and characterizing pulsar haloes w. CTA]{Detecting and characterizing pulsar haloes with the Cherenkov Telescope Array}

\author[C.~Eckner et al.]{Christopher Eckner$^{1, 2}$\thanks{Contact e-mails: \href{mailto:}{eckner@lapth.cnrs.fr, veronika.vodeb@ung.si, pierrick.martin@irap.omp.eu, gabrijela.zaharijas@ung.si, calore@lapth.cnrs.fr}}, 
Veronika Vodeb$^{3}$, 
Pierrick Martin$^{4}$,
Gabrijela Zaharijas$^{3}$, 
Francesca Calore$^{1}$  %% ... others, including ...
\\ 
$^{1}$LAPTh, CNRS,  USMB, F-74940 Annecy, France\\
$^{2}$LAPP, CNRS,  USMB, F-74940 Annecy, France\\
$^3$University of Nova Gorica, Center for Astrophysics and Cosmology, Slovenia\\
$^4$IRAP, Universit\'e de Toulouse, CNRS, CNES, F-31028 Toulouse, France\\ 
}

\pubyear{2022}

\begin{document}

\label{firstpage}
\pagerange{\pageref{firstpage}--\pageref{lastpage}}
\maketitle

\begin{abstract}
%american english
The recently identified source class of pulsar haloes may be numerous and bright enough in the TeV range to constitute a large fraction of the sources that will be observed with the Cherenkov Telescope Array (CTA). 
In this work, we quantify the prospects for detecting and characterizing pulsar haloes in observations of the projected Galactic Plane Survey (GPS), using a simple phenomenological diffusion model for individual pulsar haloes and their population in the Milky Way. 
Our ability to uncover pulsar haloes and constrain their main physical parameters in the CTA GPS is assessed in the framework of a full spatial-spectral likelihood analysis of simulated survey observations, using the most recent estimates for the instrument response function and prototypes for the science tools.
% Nice suggestion by FC to remove that 
%We derive the spectral and morphological sensitivity of the survey, first to extended sources in general, and second to the specific intensity distribution of halos. 
%From these, we quantify for which physical parameters pulsar haloes can be detected, identified, and characterized, and what fraction of the Galactic population could be accessible.
%We investigate the effect of interstellar emission and data analysis systematics on the prospects.
% ...and replace it with
%Depending on the pulsar halo model assumptions, we find that about three hundreds objects could give rise to detectable emission in the GPS survey. Yet, only a third of them could be identified through their energy-dependent morphology, and only one tenth of them would allow the derivation of strong constraints on key physical parameters like the magnitude or extent of suppressed diffusion around the pulsar. We also provide a list of known pulsars that could be hosting a detectable (Geminga-like) halo in the GPS, and assess the robustness of our findings against several systematic uncertainties. 
% Slightly modified version by PM
For a model setup representative of the halo around Geminga, we find that about three hundred objects could give rise to detectable emission in the GPS survey. Yet, only a third of them could be identified through their energy-dependent morphology, and only one-tenth of them would allow the derivation of strong constraints on key physical parameters like the magnitude or extent of suppressed diffusion around the pulsar. We also provide a list of known pulsars that could be hosting a detectable (Geminga-like) halo in the GPS and assess the robustness of our findings against several systematic uncertainties. 
\end{abstract}


\section{Introduction}\label{intro}

Over the past decade, pulsars have emerged as a significant class of objects in high-energy astrophysics, both as prominent sources in the gamma-ray sky over a broad spectral range, and as possible major contributors to the local flux of cosmic-ray leptons \cite{Evoli:2021}.

In the GeV range, the Fermi-LAT has now detected more than 270 pulsars \cite{latpulsars:2022}, which is a significant jump from the handful of objects known in the pre-Fermi era \cite{Thompson:2008}. Emission in this domain is produced by particle acceleration in the magnetosphere and/or in the striped pulsar wind beyond the light cylinder \cite{Petri:2016,Petri:2018}.

In the TeV range, three dozens sources were detected \cite{tevcat:2022} that are established or candidate pulsar wind nebulae (PWNe), $\sim1-10$\,pc-sized bubbles of hot magnetized plasma expanding beyond the pulsar wind termination shock and filled with nonthermal particles accelerated at the shock and/or in the immediate downstream turbulence \cite{Amato:2020}. Ground-based imaging Cherenkov telescopes (IACTs) like H.E.S.S., with their superior angular resolution, have played an instrumental role in the detection and study of PWNe \cite{Abdalla:2018a,Abdalla:2018b}.

The chain of transport processes shaping the evolving morphology of PWNe is yet to be fully elucidated. Detailed studies of intermediate-age objects reveal intricate emission patterns, most likely influenced by the pulsar's natal kick and the dynamics of the parent supernova remnant \cite{Abdalla:2019,Principe:2020}, limiting our ability to predict the fate of nonthermal particles and the conditions of their release into the interstellar medium (ISM). At late stages, however, when the pulsar has left its parent remnant and the nebula develops a subparsec bow-shock morphology \cite{Gaensler:2006,Bykov:2017}, particle escape may be more easily approached because the confinement volume is much smaller \citep[although the physics driving it may still be rich and subtle, with energy- and charge-dependent effects; see][]{Bucciantini:2020}.

Observations with the High-Altitude Water Cherenkov observatory (HAWC) revealed extended gamma-ray emission from regions at least $\sim 20-30$ pc in size around two middle-aged pulsars, PSR J0633+1746 and PSR B0656+14, respectively the Geminga pulsar and the pulsar in the Monogem ring \cite{Abeysekara:2017b}. This new source class seemingly consists of pulsar-powered emission from regions outside that dynamically dominated by the pulsar wind nebula and was originally dubbed TeV haloes \cite{Linden:2017}. These objects provide an opportunity to study the feedback from and escape of particles accelerated in pulsars and their nebulae. Interestingly, the Geminga and Monogem haloes can be well described with a phenomenological diffusion model, provided the value of the diffusion coefficient in the vicinity of the pulsars is much lower than the average in the Galactic medium, by two to three orders of magnitude. This implies increased level of magnetic turbulence in extended regions around pulsars, and several scenarios have been put forward to account for it: standard magnetohydrodynamical turbulence with small coherence length, possibly in the wake of the expanding forward shock parent remnant \cite{LopezCoto:2018,Fang:2019}, or self-confinement by pairs streaming from the pulsars \cite{Evoli:2018,Mukhopadhyay:2021}.

Geminga and Monogem, being among the closest middle-aged pulsars, may just be the tip of the iceberg. Since the original discovery, a number of pulsar halo candidates were proposed based on HAWC \cite{Linden:2017,Albert:2020}, H.E.S.S. \cite{DiMauro:2020}, and more recently LHAASO observations \cite{Aharonian:2021}. The phenomenon can be expected to cover a broadband spectrum and can therefore be searched for also in Fermi-LAT data, although recent attempts to detect GeV counterparts to the Geminga and Monogem haloes are so far conflicting \cite{DiMauro:2019a,Xi:2019a}. Yet, the exact physical grounds that lead to the development of pair haloes around (some) pulsars have yet to be exposed and, as a consequence, the commonness of pulsar haloes in the Galaxy is still largely unknown. Meanwhile, phenomenological models based on the assumption of suppressed diffusive transport within some distance of the pulsar provide a fairly good description of existing observations.

The detection and identification of a population of pulsar halos, including measurements of their energy-dependent morphologies over a wide spectral range, is needed for a comprehensive modeling of nonthermal particle transport in the vicinity of pulsars and the connection of high-energy electron populations near their accelerators to those present in the Galactic medium \cite{Lopez-Coto:2022igd}. Continued observations with HAWC and LHAASO, and also with H.E.S.S. pending adaptation of data analysis to very extended sources \cite{Abdalla:2021}, will no doubt be instrumental. In a few years from now, this effort will be contributed to by the Cherenkov Telescope Array (CTA), which holds special promise owing to its excellent angular resolution, broad energy coverage, and wider field of view when compared with current IACTs \cite{Acharya:2019}.  

As part of an ensemble of Key Science Projects \cite{Acharya:2019}, an extensive survey of the full Galactic Plane will be performed with CTA, using both the Southern and Northern CTA observatories, resulting in unprecedented exposure, both in depth and footprint (the so-called Galactic plane survey, GPS). This effort will be accompanied by the production of catalogs \cite{Remy:2022} that will seed deeper investigations and provide opportunities for serendipitous discoveries, including TeV halos.

In this work, we aim to assess the potential for the upcoming CTA experiment to constrain the current phenomenological halo models, based on the GPS observations. We determine the sensitivity of the survey to the emission spectrum and morphology predicted for individual haloes with different model setups. This is achieved in the framework of a full spatial-spectral likelihood analysis, using the most recent estimates for the instrument response function and prototypes for the science tools, and including an investigation of the effect of interstellar emission and data analysis systematics. We then frame the results in the context of a population synthesis, to estimate the fraction of the population that could be within the reach of CTA, and we provide a list of promising targets among known pulsars.

%CTA is the next generation ground-based observatory for gamma-ray astronomy at very-high energies in the range between 20 GeV and 300 TeV. CTA will consist of an array of imaging atmospheric Cherenkov telescopes (IACTs) at two different locations, one on the Southern and another at a Northern hemisphere, enabling it to observe the whole sky.  

%Detecting the pulsar haloes also with CTA and understanding better their properties can contribute to answering many open questions, such as understanding the particle acceleration processes and diffusion processes that happen in our Galaxy. It can bring more understanding of the origin of electrons and positrons, as constrains can be put on the origin of the positron flux on Earth, as well as to put constraints on the values of diffusion coefficients. As DM sub-halos can hide among {\it extended} gamma-ray sources (cite), TeV pulsar halo identification can contribute significantly to reducing the backgrounds in the indirect DM searches.\\

%In the analysis, we focus both on the planned GPS observation, but consider also an impact of dedicated pointing observations of selected PSR halo candidates. 
%\gabi{do we do the later?}
% PIMA: We decided not to, so leave it out
%\red{@Francesca, Add something  about dark subhalos in the context of extended sources, if we decide to go for that.}

The present manuscript is organized as follows: in Sec.~\ref{model}, we present the different components that compose our model for the gamma-ray sky at TeV energies;
in Sec.~\ref{data}, we describe the procedure adopted to simulate the survey's observations, 
as well as the analysis framework; in Sec.~\ref{sensitivity}, we illustrate the results of the CTA sensitivity to 
pulsar halos, under different analysis' assumptions; we discuss the physics prospects and implications for the study of the pulsar haloes Galactic population in 
Sec.~\ref{phy}; finally, we draw our conclusions in Sec.~\ref{conclusion}.

%%%%%%%%%%%%%%%%%%
\section{Sky model}\label{model}
%%%%%%%%%%%%%%%%%%

In this section, we introduce the astrophysical emission components considered in our simulations and analyses of CTA observations. We first describe the phenomenological halo model and in particular the single representative model setup that is later used in our assessment for detectability of individual halos. We then describe how such a model was used in another work by one of the authors to produce a synthetic halo population for the whole Galaxy, which we will use to infer the fraction of the sources that could be within reach of CTA in the context of GPS observations. At last, we present different models for the astrophysical background components considered in this work, namely the interstellar emission from the large-scale population of cosmic rays (CRs) of the Galaxy.

%%%%%%%%%%%%%%%%%
\subsection{Individual halo model}\label{model:halo}
%%%%%%%%%%%%%%%%%

The individual halo model and the reference parameter setups used in this work were introduced in \citet{Martin:2022} and derived mostly from the formalism exposed in \citet{Tang:2019}. We refer the reader to these publications for details and just summarize here the salient features of the model.

A few tens of kyr after pulsar birth, when, as a result of its natal kick, the pulsar exits its parent remnant or the original pulsar wind nebula, electron-positron pairs accelerated in the now bow-shock pulsar wind nebula are released into the surrounding medium. This happens on short time scales, mainly due to the small subparsec size of the object. Particles then diffuse away isotropically in a medium characterized by a two-zone concentric structure for diffusion properties. An outer region is representative of the large-scale average ISM in the Galactic plane, while the inner region with a radial extent of a few tens of pc has diffusion suppressed by factors of a few tens to a few hundreds as a result of an unspecified physical mechanism (for instance preexisting fluid turbulence with small coherence length or self-confinement of the streaming pairs). Along their propagation, particles lose energy and radiate via the synchrotron and inverse-Compton scattering processes, in magnetic and radiation fields assumed to be typical of the ISM.

In the absence of a consistent physical description for the very origin of particle confinement around (some) pulsars, the conditions for suppressed diffusion are assumed to be static and the suppressed diffusion coefficient follows a power-law rigidity dependence with index 1/3, applicable for scattering in magnetic turbulence with a Kolmogorov spectrum. Following \citet{Martin:2022} and \citet{Martin:2022b}, we considered different possible scenarios for the diffusive halos: three different radial extensions of the suppressed diffusion zone (30, 50, or 80 pc), and two levels of diffusion suppression with respect to large-scale interstellar conditions (by a factor 500, as required by observations of J0633+1746, or by a factor 50, in agreement with observations of B0656+14 within uncertainties as shown in \citet{Martin:2022}; in practice this translates into diffusion coefficients of $4 \times 10^{27}$ and $4 \times 10^{28}$\dunit at 100\tev, respectively).

We consider, as baseline model, a halo with the following properties:
\begin{enumerate}
\item A pulsar with a current age of 200\,kyr, spin-down power of $10^{35}$\punit, and spin-down evolution with a braking index of 3;
\item Particle injection is assumed to start 60\,kyr after pulsar birth and to last until the current pulsar age with a constant injection efficiency initially set at 100\%;
\item Injected particles have a constant broken power-law spectrum from 1\gev up to a cutoff at 1\pev, with indices 1.5 and 2.4 below and above a break at 100\gev, respectively;
\item A suppressed diffusion region size of 50\,pc and a level of diffusion suppression in this region by a factor 500;
\item The interstellar magnetic and radiation fields are those corresponding to a position at (R, z)=(4500\,pc, 0\,pc) in the Galactic population synthesis of \citet{Martin:2022b};
\item The effects of proper motion on the halo morphology are neglected, which is justified by the fact that very-high-energy signatures of haloes are little sensitive to it \cite{Zhang:2020a}.
\end{enumerate}
The 60\,kyr start time for particle injection is the typical time at which a pulsar enters the bow-shock phase, during which accelerated particles are expected to escape the nebula quite rapidly. It corresponds to the time when the pulsar exits its parent remnant, or the original nebula at its center, owing to its natal kick \cite[see the discussion in][]{Martin:2022}. The 1\pev cutoff energy exceeds by a factor $\sim2$ the highest particle energy that can be reached from the maximum potential drop under the assumption of ideal magnetohydrodynamics \cite{DeOnaWilhelmi:2022}. In a population study perspective, the exact maximum energy in the particle spectrum has a limited influence on the prospects as the CTA sensitivity to haloes is mostly found at gamma-ray energies below a few tens of TeV (see Sect. \ref{sensitivity}), in a range where the signal is mostly contributed to by $\lesssim$100\tev particles inverse-Compton scattering infrared photon fields (see also the discussion in Sect. 2.1 of \cite{Martin:2022}).

Variants of the above model setup were computed for alternative suppressed diffusion region sizes (30 or 80 instead of 50\,pc) and level of diffusion suppression (50 instead of 500). Three-dimensional model cubes for these individual halo model setups were computed over a range of distances from 1 to 15\,kpc, at the reference coordinates $(l,b) = (-10^{\circ}, 0^{\circ})$ used in our assessment of the prospects for detection and study (see Sect. \ref{data:ana}). We note that putting this test halo model at various distances from 1 to 15\,kpc for the reference position causes an inconsistency because the interstellar magnetic and radiation fields used by default are not appropriate to all locations considered. Yet, we decided not to change distance and environmental conditions at the same time to avoid mixing too many different effects in the detectability trends. 

When deriving prospects for detectability in the following, a given halo model setup will be renormalized to the minimum value required to achieve a given scientific goal (for instance simple detection with $TS=25$, or detection as an energy-dependent extended source). This is done via the injection luminosity $L_{\rm inj}$ parameter, which is the product of injection efficiency and present-day spin-down power. 

%\gabi{Since the extension of the halo is degenerate with  other parameter (notably the electron  injection index) in terms of  the predicted gamma-ray emission, we consider the 30pc  extension as  the benchmark assumption and explore other values only in dedicated plots.}

%%%%%%%%%%%%%%%%%
\subsection{Halo populations model}\label{model:pop}
%%%%%%%%%%%%%%%%%

The individual halo model setups described above were used to assess the conditions under which a typical halo can be observed in the CTA GPS. This means, for instance, evaluating the particle injection power required for a halo to be detected as an extended source with an energy-dependent morphology. Such constraints determined on representative halo models are then compared to the properties of a Galactic synthetic population to infer the number of objects that could be accessible to the survey. For that purpose, we used a halo population synthesis that was introduced in \citet{Martin:2022b} and we just summarize here its main features. 

The full population model starts with the generation of a synthetic population of young pulsars, with random selection of positions, powers, natal kick velocities, and ages. Each pulsar initially feeds a PWN until the time when it exits its nebula, which marks the beginning of the halo phase. haloes are modeled as described in the previous section and all objects in a given population share the same properties for the suppressed diffusion region. Conversely, the properties of injected particles were randomly selected from statistical distributions of power-law indices above the break energy, cutoff energies, and injection efficiencies. The parameters of the entire population of sources, haloes and PWNe (and also SNRs), were calibrated so that the flux distribution of mock objects in the TeV range match that of known sources. A typical realization of the population includes about 2600 objects, with ages from 20 to 400\,kyr,  and 1--10\tev luminosities in $10^{30}\mbox{--}10^{34}$\punit.

We emphasize here a caveat in the approach defined above: when assessing the detectable fraction of the population in Sect. \ref{phy:popu}, we will compare haloes from a synthetic population comprising a large variety of conditions (pulsar age, magnetic and radiation field intensities, particle injection start time and spectrum,... ) with detectability criteria that are strictly valid for only one set of halo model parameters. Yet, assessing the detectability for a large number of halo parameters combinations would have been prohibitive computationally speaking, hence our choice of the above approach. For that reason, the prospects at the population level, introduced in Sect. \ref{phy:popu}, should be taken as approximations. 

%%%%%%%%%%%%%%%%%
\subsection{Large scale diffuse backgrounds}
%%%%%%%%%%%%%%%%%

In addition to the emission from pulsar halos, and the instrumental background that will be introduced in the next section, we included in our analyses a model for large-scale interstellar emission (IE) from interactions of the Galactic population of CRs with the ISM. Such a component can have important effects on the individual detectability of extended sources like halos.

The IE runs predominantly along the Galactic plane and has been exquisitely mapped with the Fermi-LAT at GeV energies, where it is the brightest emission component in the sky (see \cite{Fermi-LAT:2012edv} for a review). In the TeV range, however, the properties of this source are less solidly established, in part because of difficulties to detect emission components extending over angular scales larger than the field of view with IACTs. There has been, however, recent progress in this domain thanks to the advent of ever-sensitive water-Cherenkov detectors like MILAGRO, HAWC, or LHAASO, owing to their very large effective area and instantaneous field of view and high duty cycles.

In order to model this component we take advantage of a recent study \cite{Luque:2022buq} based on available GeV to PeV 
gamma-ray data (from Fermi-LAT, Tibet AS$\gamma$, LHAASO and ARGO-YBJ), together with local charged CR measurements (from AMS-02, DAMPE, CALET, ATIC-2, CREAM-III, and NUCLEON). Modeling the IE is achieved within two physical frameworks: in the so-called ``Base'' models the diffusion coefficient is assumed to be constant throughout the Galaxy,
while it is allowed to vary radially in the ``$\gamma$-optimized'' models. Both sets of models are 
further divided in ``Min'' and ``Max'' setups in order to reflect uncertainties in the CR proton and Helium source spectra, 
see Ref.~\cite{Luque:2022buq} for more details. We chose the ``Base Max'' setup as our benchmark model, but explored in Sect. \ref{sec:robustness} the impact on the derived sensitivities to haloes of using ``$\gamma$-optimized Min'' model instead.

For consistency, we also consider a model adopted in a reference prospect study of the CTA GPS (the so called 'GPS IEM' in Ref. \cite{Dundovic:2021ryb}). This model is tuned to direct CR measurements near the Earth, but agnostic to current gamma-ray measurements that indicate higher and harder CR spectra in central regions of the Galaxy. The `GPS IEM' model therefore represents a minimal contribution expected from IE at TeV energies. 

%%%%%%%%%%%%%%%%%
\section{Data simulations and analysis}\label{data}
%%%%%%%%%%%%%%%%%

%%%%%%%%%%%%%
\subsection{Simulation of survey observations} \label{data:sim}
%%%%%%%%%%%%%%

The CTA GPS will provide a view of the Galactic Plane at TeV energies with unprecedented depth, angular resolution, and spectral coverage. 
The observational campaign will consist of a short-term program, with 480\,h of observing time allocated over the first two years, and a long-term program, with 1140\,h of observing time allocated over the following eight years \cite{Acharya:2019}. Regarding the practical implementation of the survey in our simulations, we followed the same approach as in Ref. \cite{Remy:2022}. Observations are distributed along the plane following a double-row pointing pattern and data are taken by both the North and South arrays, according to a preliminary schedule that reduces contamination by Moon light and minimizes zenith angle in each observation, so as to optimize the performances of the instrument \cite{Remy:2022}.
%\gabi{Which times we use, the full 1140h?}
% Pierrick: Yes
% \gabi{Shall we put the GPS in context - mention HGPS, HAWC etc? @Pierrick} 
% Pierrick: I do not think so, this does not add relevant information to what we describe here? We can refer to other papers instead.
%The previous surveys of the plane, performed by the H.E.S.S. already resulted in a number of discoveries including XX and detection of the cumulative  diffuse emission at TeV energies.  

\begin{figure}
\centering
\includegraphics[width=0.45\textwidth]{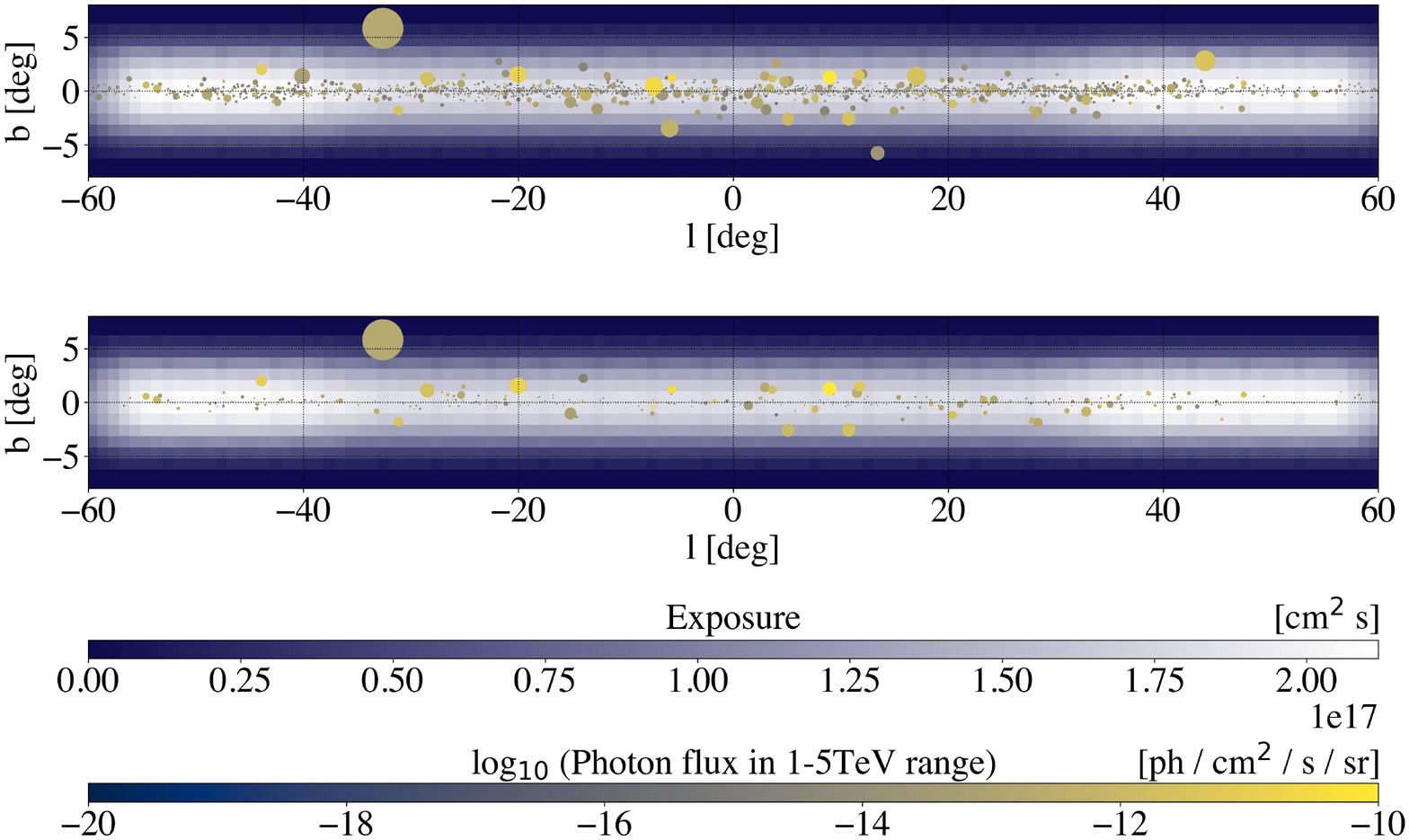}
\caption{The $0.1-100$\tev exposure for the planned GPS observations overlaid with one realization of a synthetic pulsar halo population. {\it Top:} All simulated sources. {\it Bottom:} Only resolved sources (see Sec. \ref{phy:popu}).  \label{fig:GPSexp}}

\end{figure}

Due to the different configurations of the North and South arrays, and because of the particular importance of specific regions in our Galaxy, the planned exposure in the GPS varies along the plane and has been divided into five segments: Inner Galaxy, Cygnus/Perseus, Anticenter, and two chunks for Vela/Carina \cite{Acharya:2019}. The anticipated exposure is illustrated in Fig. \ref{fig:GPSexp} for the Inner Galaxy within longitudes $-60^{\circ} < l < 60^{\circ}$ and latitudes $-3^{\circ} < b < 3^{\circ}$ (for full exposure see Appendix \ref{app:fullGPS}). This region will be observed with the deepest exposure as it harbors the largest density of sources. Since this is also the region where the majority of young pulsars, hence pulsar halos, are expected, we use as reference position for our study the coordinates $(l,b) = (-10^{\circ}, 0^{\circ})$. On the other hand, this may appear as a conservative choice, since the Inner Galaxy is where IE is the most intense at this position.

We used the most recent instrument response functions (IRFs) for the observatory\footnote{see \url{http://www.cta-observatory.org/science/cta-performance/} for more details}, labeled \texttt{prod5}, for the envisaged configurations of the North and South sites. The respective IRF data files are publicly available at \cite{cherenkov_telescope_array_observatory_2021_5499840}. We used by default the ``alpha'' configuration, which refers to the arrays that will be built in an initial phase of the project, with a realistic number of telescopes given the currently secured budget. The spatial distribution of telescopes in this configuration has been optimized for performance across a series of science goals.

% OMEGA CONFIGURATION? 
%A potential extension of the instrument towards the configuration that was initially planned, dubbed the ``omega'' configuration, may follow and we discuss in appendix the impact it may have on some key results of our study. \gabi{Shall we do this? where to show this line, redo all plots or  only on the population study plot maybe?}

In practice, we selected the set of IRFs corresponding to event reconstruction quality and background cuts optimized on the basis of Monte Carlo simulations of 50\,h of observations, which is appropriate for studies of extended sources. By background, we refer here to air showers triggered by CR particles entering the atmosphere that can be misidentified as gamma rays in the event reconstruction process. These actually make up the largest fraction of the events detected by ground-based observatories, and this will remain true for the CTA. The misidentified CR component is modeled from extensive Monte-Carlo simulations of CR air showers, detection of the associated Cherenkov radiation, and subsequent event reconstruction. The corresponding rate and distribution of misidentified CR events is provided with the suite of IRFs for each given set of observing conditions (duration, zenith angle,...). There are however significant uncertainties or biases in this predicted component, in normalization, spectrum, and spatial distribution within the field of view, such that, in an actual data analysis, it is generally needed to fit it the data (simultaneously to the models for the emission of astrophysical sources). In the spatial-spectral likelihood analysis introduced below, the normalization of the CR background is therefore left free, which is a minimum assumption to cover the uncertainties in this component. Last, simulations and analyses were performed using the publicly available gamma-ray analysis software \texttt{ctools}\footnote{\url{http://cta.irap.omp.eu/ctools/}}.
	
 %Since the CTA IRFs are furthermore optimised regarding the maximum zenith angle at which a target is observed, we choose the best-suited IRF file according to the expected zenith angle (given the location of CTA North and South on the Earth) of the pointing position's centre stated in this list. 
	
%%%%%%%%%%%%%%%%%%%%%
\subsection{Analysis framework}
%%%%%%%%%%%%%%%%%%%%%
\label{data:ana}

To evaluate our capability to detect and study pulsar haloes with CTA, we adopt the statistical inference framework of Ref.~\cite{CTA:2020qlo} -- there employed in the context of indirect searches for a continuum gamma-ray signal from dark matter pair-annihilation in the Galactic center region with CTA -- and adapt it to our needs. 

Source characterization in simulated GPS observations is achieved by maximum likelihood estimation of the parameters of an emission model for some region of interest. We perform a three-dimensional likelihood analysis for binned data and Poisson statistics. Unless otherwise stated, the analysis is performed under the following conditions:
\begin{enumerate}
\item Region of interest (ROI) centered on $(l,b) = (-10^{\circ}, 0^{\circ})$, with $6^{\circ} \times 6^{\circ}$ size, aligned on Galactic coordinates, with $0.02^{\circ} \times 0.02^{\circ}$ pixel size.
\item Energy range from 0.1 to 100\tev, binned logarithmically with 15 bins per decade\footnote{We use 15 logarithmically spaced energy bins to ensure that the IRFs are properly sampled. This finer binning is summed into larger energy bins before performing the analysis.}. Energy dispersion is not used.
\end{enumerate}

Our model for the signal in the ROI is based on templates of the expected instrumental background $\{B_l\}_{l\in L}$ and astrophysical signal components $\{S_k\}_{k\in K}$ with $L$ and $K$ being index sets for each series of templates (for instance different background noise templates for the different pointings in a data set, and different emission components for the region of the Galactic plane under scrutiny). %\ce{This version reads like IE and pulsar halo are part of the astrophysical signal. I would define the IE as part of the background components $B$ to be clearer.}\pima{Yes, IE formally goes in astrophysical signals, although I can understand Christophers' suggestion to put it in a category with every other source that pollutes the signal of interest to us. They are of different nature, however, and are usually handled differently in realistic data analysis.}\gabi{@CE; revoke the  changes}

The Poisson likelihood function reads
\begin{equation}
\label{eq:poisson_likelihood}
\mathcal{L\!}\left(\left.\bm{\mu}\right|\bm{n}\right)=\prod_{i,j}  \frac{\mu_{ij}^{n_{ij}}}{\left(n_{ij}\right)!}e^{-\mu_{ij}}\mathrm{,}
\end{equation}
where $\bm{\mu}$ denotes the user-defined model cube describing the signal whereas $\bm{n}$ refers to the experimental data taken by CTA, here realized via simulated ``mock" observations. The index $i$ labels the energy bins while the index $j$ enumerates the spatial pixels of our templates. Our model is a linear combination of templates describing distinct contributions to the signal received by the instrument. In the most general form
\begin{equation}
\label{eq:model_gen_eq}
%\bm{\mu}= \sum_{k\in K}\theta^{S_k} \bm{S}_k + \sum_{l\in L}\sum_{i}\theta_{i}^{B_l}B_{l,i}\rm{,}\label{eq:model_eq}
\bm{\mu}= \sum_{k\in K} \bm{S}_k (\theta^S_k) + \sum_{l\in L} \bm{B}_l (\theta^B_l) \mathrm{,}
\end{equation}
where the vector sets $\{\theta^S_k\}_{k \in K}$ and $\{\theta^B_l\}_{l \in L}$ are parameters affecting the spectral and angular dependence of the astrophysical signal and instrumental background noise templates, respectively. In practice, however, most of the cases we handle involve a global renormalization of predefined astrophysical signal and instrumental noise templates, possibly with an energy dependence in the case of the instrumental noise, such that the model prediction reads
\begin{equation}
\label{eq:model_spe_eq}
%\bm{\mu}= \sum_{k\in K}\theta^{S_k} \bm{S}_k + \sum_{l\in L}\sum_{i}\theta_{i}^{B_l}B_{l,i}\rm{,}\label{eq:model_eq}
\mu_{ij}= \sum_{k\in K} \theta^S_k S_{k,ij} + \sum_{l\in L} f^B_i(\theta^B_l)B_{l,ij} \mathrm{,}
\end{equation}
The presence of a spectral correction $f^B_i(\theta^B_l)$ in the case of instrumental background is warranted by our imperfect knowledge of the instrumental response in realistic conditions, and therefore the possibility of a cross-talk between components when testing a given extended astrophysical model.

%By introducing this distinction between the modeling of signal and background emission, it is possible to account for spectral fluctuations present in the observational data or a potential mismodeling of some aspects of the employed emission models. The latter aspect is especially relevant when dealing with experimental data because theoretical or data-driven models are never expected to be a perfect representation of reality.

Our ability to discriminate between alternative hypotheses regarding the composition of the measured signal within a particular ROI is assessed via the log-likelihood ratio test statistic (TS), which quantifies the significance of the improvement provided by a test hypothesis over a null hypothesis when compared to the data. In general, the TS reads
\begin{equation} 
\label{eq:ts} 
\textrm{TS} = 2 \left( \ln \left[ \frac{\mathcal{L} \left( \bm{\mu} ( \{ \hat{\theta}^{S_{\rm test}}_k \}, \{ \hat{\theta}^B_l \}) | \bm{n} \right)}{\mathcal{L} \left( \bm{\mu} (\{ \hat{\theta}^{S_{\rm null}}_k \}, \{ \hat{\theta}^B_l \}) | \bm{n} \right)} \right] \right) \mathrm{,}
\end{equation}
where hatted quantities refer to the best-fit values for all model parameters (astrophysical signal and instrumental background noise) obtained after a maximum likelihood fit to the mock data using Eq.~\ref{eq:poisson_likelihood}. By profiling over the background parameters, we treat them as nuisance parameters. %\ce{I have slightly changed the equation above because it did not reflect the profiling over nuisance parameters (which is what Veronika and I have done in practice). In addition, the signal parameters $\theta^{S_{\rm test}}_k$ should be free in the alternative hypothesis and not fixed to a maximum likelihood fit.} 
%\pima{I am not sure to understand what you mean here by profiling. TS is the ratio of the log-likelihoods of the two hypotheses, resulting from optimization of all free parameters in each hypothesis. The hatted quantities actually meant these vectors or model parameters after optimization, and these vectors can contain fixed or free parameters. To me, the min() is redundant with the hat, because hat is what you get after optimization. Did I miss anything ?} \ce{No, you did not miss anything but I was looking at the problem from another perspective. I did not think about $S_k$ as containing both pulsar halo and other astrophysical background components. I also like my TS to be a true function of the signal parameters (halo normalization). Using best-fit values is a totally valid way of stating this equation, I was just confused since I did not think about it in full detail.} \gabi{@CE}
Examples of hypotheses that will be tested in the following include simple detection of a model component (the null hypothesis is a fit without that model component), detection of the pulsar halo signal beyond a certain physical radius (the null hypothesis is a fit with a model clipped to zero beyond that radius), or the detection of the halo energy-dependent morphology (the null hypothesis is a fit with an energy-independent model such as a two-dimensional Gaussian).

%To derive a forecast of CTA's detection sensitivity regarding pulsar halos, we implement the following version of the log-likelihood ratio test statistic \cite{Cowan:2010js}:
%\begin{widetext}
%\begin{equation} 
%\label{eq:TS_discovery} 
%\textrm{TS}_{\mathrm{discovery}}\!\left(\theta^{S_k}\right) = \begin{cases} -2\min_{\{\theta_i^{B_l}\}}\left(\ln\!\left[\frac{\mathcal{L}\!\left(\left.\bm{\mu}(\{\theta^{S_k}\} = 0,\,\theta_i^{B_l}) \right|\bm{n}\right)}{\mathcal{L}\!\left(\left.\bm{\hat{\mu}}\right|\bm{n}\right)}\right]\right)\, & \forall k \in K: \hat \theta^{S_k} \geq 0  \\ 0 &  \exists k\in K: \hat\theta^{S_k} < 0\rm{,}
%\end{cases} 
%\end{equation}
%\end{widetext}
%where quantities denoted by $\hat\cdot$ refer to the best-fit values for all normalization parameters (signal and background) obtained after a maximum likelihood fit to the mock data using Eq.~\ref{eq:poisson_likelihood}. By profiling over the background parameters, we treat them as nuisance parameters. The construction in Eq.~\ref{eq:TS_discovery} thus probes the preference for a model that includes the assumed signal components over a model that only contains the background templates. 

%%%%%%%%%%%%%%%%%%%%%
\subsection{Detectability estimates}
%%%%%%%%%%%%%%%%%%%%%
\label{data:det}

In the simple case where the detection of astrophysical components is tested, Eq. \ref{eq:ts} reduces to 
\begin{equation} 
\label{eq:tsdet} 
\textrm{TS}_{\mathrm{det}} = 2 \left( \ln \left[ \frac{\mathcal{L} \left( \bm{\mu} ( \{ \hat{\theta}^S_k \}, \{ \hat{\theta}^B_l \}) | \bm{n} \right)}{\mathcal{L} \left( \bm{\mu} (\{ \theta^S_k=0 \}, \{ \hat{\theta}^B_l \}) | \bm{n} \right)} \right] \right) \mathrm{.}
\end{equation}
The significance of the detection is given by the value of the test statistic and depends on the number of signal components. It can be shown that, under a set of regularity conditions, statistical fluctuations in the absence of signal components in the data result in $\textrm{TS}_{\mathrm{det}}$ being distributed according to a non-central $\chi^2$-distribution with $K$ degrees of freedom \cite{Cowan:2010js}. In what follows, we usually consider the case where we only have a single signal template. In this particular case, we find that the distribution $p(\textrm{TS}_{\mathrm{det}})$ reads:
\begin{equation}
\label{eq:proba} 
    p(\textrm{TS}_{\mathrm{det}}) = \frac{1}{2} \left( \delta(\textrm{TS}_{\mathrm{det}}) + \frac{1}{\sqrt{2\pi\textrm{TS}_{\mathrm{det}}}} e^{-\frac{\textrm{TS}_{\mathrm{det}}}{2}} \right)\rm{,}
\end{equation}
where $\delta$ is the Dirac $\delta$-distribution. The entire probability distribution is sometimes called a half-$\chi^2$-distribution with one degree of freedom. The test statistic value corresponding to a signal detection at the $5\sigma$ level is $\sim 25$.

%%%%%%%%%%%%%%%%%%%%%
\subsection{Sensitivity analysis}
%%%%%%%%%%%%%%%%%%%%%
\label{data:sens}

\textbf{Spectral sensitivity:} Differential spectral sensitivity was computed over the full energy range for 5 bins per decade, which implies casting the original binning of our model templates on a coarser grid (see Sect. \ref{data:ana}). The sensitivity to a given source of interest in each independent energy bin is computed as the flux level yielding a detection with a log-likelihood ratio $TS = 25$ in that bin alone\footnote{We do not implement any additional constraint such as requiring at least ten detected gamma rays per energy bin and a signal to background ratio of at least 1/20, as done to obtain the differential sensitivity shown in the CTAO performance page for \texttt{prod5} IRFs.}. Mock observations are simulated from an emission model containing the instrumental background noise and astrophysical signal components (the source of interest and possibly another component such as the IE model). For a given normalization of the component for the source of interest and in a given energy bin, a $TS$ is computed from the likelihoods obtained in two fits: a fit for the test hypothesis in which the emission model consists in all components used in the data simulation, and a fit for the null hypothesis in which the component for the source of interest was removed. The normalization of the source of interest is iteratively adjusted and the observation simulation and data analysis sequence is repeated until the $TS$ converges towards a value of $25$ in that energy bin for that source of interest.

\noindent\textbf{Model-independent angular sensitivity:} Angular sensitivity was computed over three energy bands ($0.1-1$, $1-10$ and $10-100$\tev), which implies reducing our original binning scheme to just three two-dimensional maps. We used a similar approach as for the spectral sensitivity analysis and searched for the normalizations of a set of $N$ concentric uniform-brightness annuli with mean radii $\{r_i\}$ and widths $\{\Delta r_i\}$, with $i=1...N$, such that each annulus is individually detected with $TS = 25$ in a given energy range. We assumed a fixed power-law spectrum with photon index 2.0 for each annulus and typically used $\Delta r_i = 0.2^{\circ}$ and angular distances from the center of the ROI $r_i = 0^{\circ}$...$3^{\circ}$. Mock observations are simulated from an emission model containing the instrumental background noise, an interstellar emission template, and one given annulus. For a given brightness of the annulus, a $TS$ is computed in each energy range from the likelihoods obtained in two fits: a fit for the test hypothesis in which the emission model consists in all components used in the data simulation, all with free normalizations, and a fit for the null hypothesis in which only the instrumental background noise and interstellar emission templates are fitted. The brightness of the annulus is iteratively adjusted and the observation simulation and data analysis sequence is repeated until the $TS$ converges towards a value of $25$ for that annulus in that energy range. The process is performed for all annuli independently in all three energy ranges.

\subsection{Angular decomposition of a pulsar halo signal}

We perform an angular decomposition of individual pulsar haloes modeled as introduced in Sec.~\ref{model:halo}. We start from a given mock data set $\bm{n}$ derived from templates for the instrumental background, IE, and a pulsar halo for a particular choice of injection power, diffusion region size and distance to the Sun, with all components included at their nominal values according to the utilized flux models, i.e.~$\theta_k^S\equiv1, \theta_l^B\equiv1\;\forall k\in K,l\in L$. We conduct successive fits of a growing number of concentric annuli of fixed width $\Delta r$, obtained by truncating the original pulsar halo model. The successive fit rationale is as follows: starting with the central annulus -- a disc of radius $\Delta r$ --  as the only pulsar halo component in the model $\bm{\mu}$, we evaluate Eq.~\ref{eq:tsdet}, which either yields $\mathrm{TS}_{\mathrm{det}} < 25$ or $\mathrm{TS}_{\mathrm{det}} > 25$. In the $\mathrm{TS}_{\mathrm{det}} < 25$ case, we increase the annulus width until we either get $\mathrm{TS}_{\mathrm{det}} > 25$ and continue the iterative fit, or a maximum allowed width is reached and we stop the decomposition. In the $\mathrm{TS}_{\mathrm{det}} > 25$ case, we continue by treating the previously considered annulus as an astrophysical components and adding the next annulus to the model. We evaluate again Eq.~\ref{eq:tsdet} with the model now containing the additional second annulus and determine whether this new component is significantly detected, i.e.~the model including the two annuli is statistically favored over that containing only the first annulus. In case of a positive result, i.e.~$\mathrm{TS}_{\mathrm{det}} > 25$ is obtained for the additional annulus, we iterate this procedure until adding another annulus to the fit does not result in any significant improvement. Eventually, the best-fit parameters and errors for all significant annuli in the last iteration step are taken as the recovered angular decomposition of the input signal.

%%%%%%%%%%%%%%%%%%%%%
\subsection{Treatment of systematics}
%%%%%%%%%%%%%%%%%%%%%
\label{data:syst}

The Poisson likelihood function in Eq.~\ref{eq:poisson_likelihood} incorporates knowledge about the probabilistic nature of the process to detect gamma rays and makes it possible to deal with the statistical uncertainty of a measurement arising due to its Poisson nature. However, CTA is going to observe the Galactic plane for an extensive period of time and large amounts of data will be accumulated, such that we will transition from a statistics-limited regime to a systematics-dominated one in a certain energy range. This means that, after this transition, taking additional data does not improve the sensitivity of the instrument anymore since systematic effects and uncertainties become relevant and may bias the search for a certain signal. To characterize the expected impact of systematic uncertainties, we introduce an effective Poisson likelihood function (see \cite{Silverwood:2014yza} for a more detailed explanation)
\begin{equation}
\label{eq:likelihood_sys}
    \mathcal{L}\!\left(\bm{\mu},\bm{\alpha}\left| \bm{n}\right.\right)=\prod_{i,j}\frac{(\mu_{ij}\alpha_{ij})^{n_{ij}}}{\sqrt{2\pi}\sigma_{\alpha}n_{ij}!}e^{-\mu_{ij}\alpha_{ij}}e^{\frac{(1-\alpha_{ij})^{2}}{2\sigma_{\alpha}^{2}}}\mathrm{,}
\end{equation}
which modifies the standard likelihood function through the addition of nuisance parameters $\bm{\alpha}$ centered around one with variance $\sigma_{\alpha}^2$ per pixel and energy bin of the binned model. These nuisance parameters can be seen as Gaussian noise that affects each pixel independently of the others leading to up- or downward fluctuations. However, this effective treatment of systematic uncertainty can only cover such sources of uncertainty that enter linearly in the calculation of the number of expected gamma rays, for instance, the effective area of CTA. Incorporating non-linear effects is much more involved. It requires the use of Monte Carlo simulations to estimate the likelihood landscape of such contributions, which goes beyond the scope of this work and analyses based on simulated mock data in general. 

To derive detection sensitivities in the framework of Eq.~\ref{eq:likelihood_sys}, we add the parameters $\bm{\alpha}$ to the set of background parameters that are profiled over in Eq.~\ref{eq:tsdet}. We remark that this implementation of systematic errors introduces a certain dependence on the binning of the utilized templates in terms of a characteristic length scale equal to the spatial bin size. Since this prescription assumes all pixels to be affected independently of each other by the systematic effects, it is a statement about the statistical independence of the pixels, which is a priori not guaranteed. A more detailed discussion of this limitation can be found in Refs.~\cite{Bartels:2017dpb, CTA:2020qlo}. We discuss the impact of systematic uncertainties for the detection and characterization of pulsar haloes in Sec.~\ref{sec:robustness}.

%\pima{Here describe the IRFs used and the configurations/layouts they correspond to. We go for prod5 alpha as baseline, and can present prod5 omega/full-array as variant for some key result. Note that the GPS group will use prod3b for the full array because they started with that.}

%\pima{describe IE models. We use the same as used in the ongoing GPS analysis/paper. The so-called BASE and GAMMA models are outdated, and the GPS group moved to a MINIMAL model as baseline and GAMMA+ as variant. SO we should use these too, with MINIMAL as baseline.}
	
%%%%%%%%%%%%%%%%%
\section{CTA sensitivity to pulsar halos}\label{sensitivity}
%%%%%%%%%%%%%%%%%%%
	
\begin{figure*}
\begin{centering}
\includegraphics[width=0.49\linewidth]{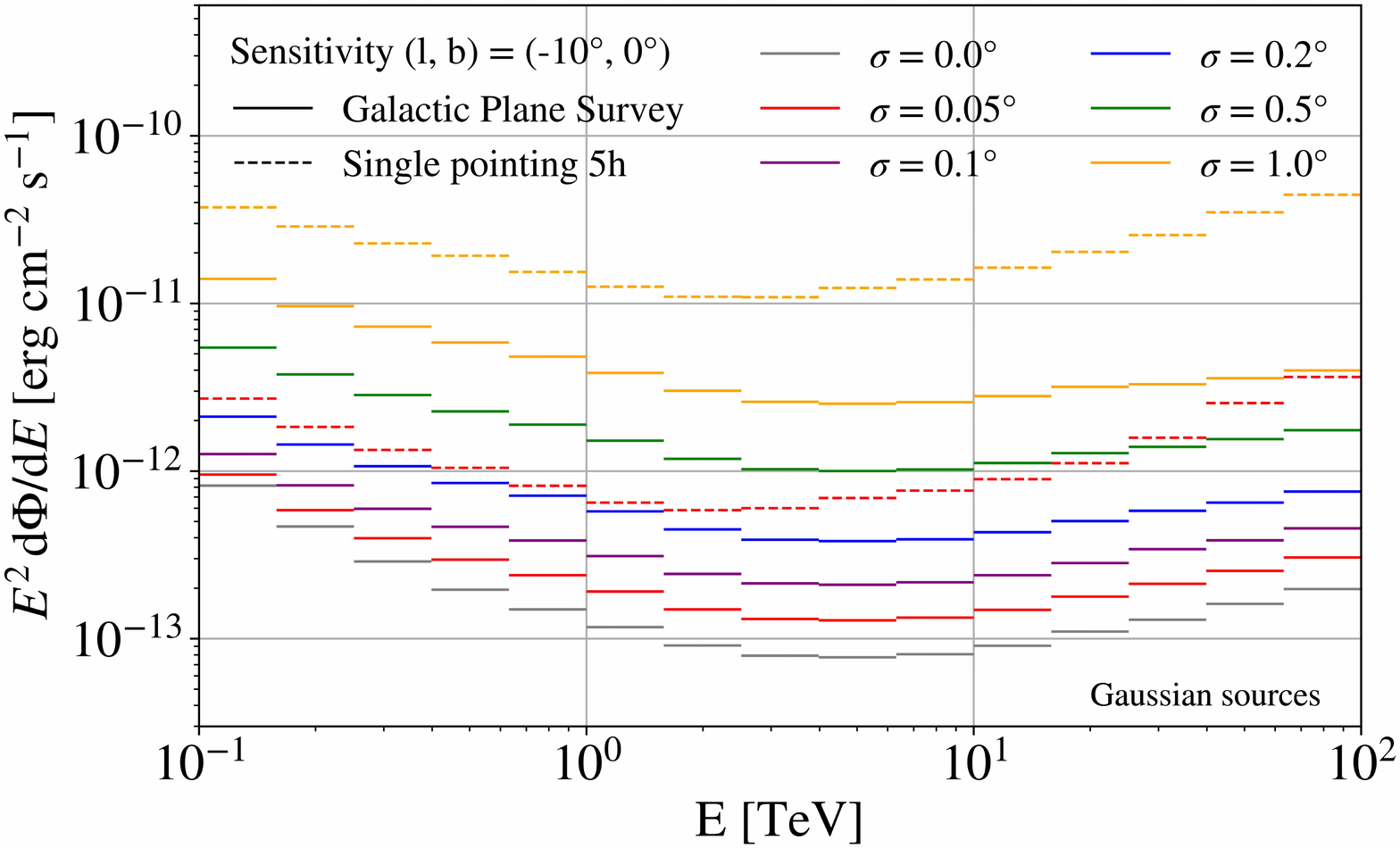}
\includegraphics[width=0.49\linewidth]{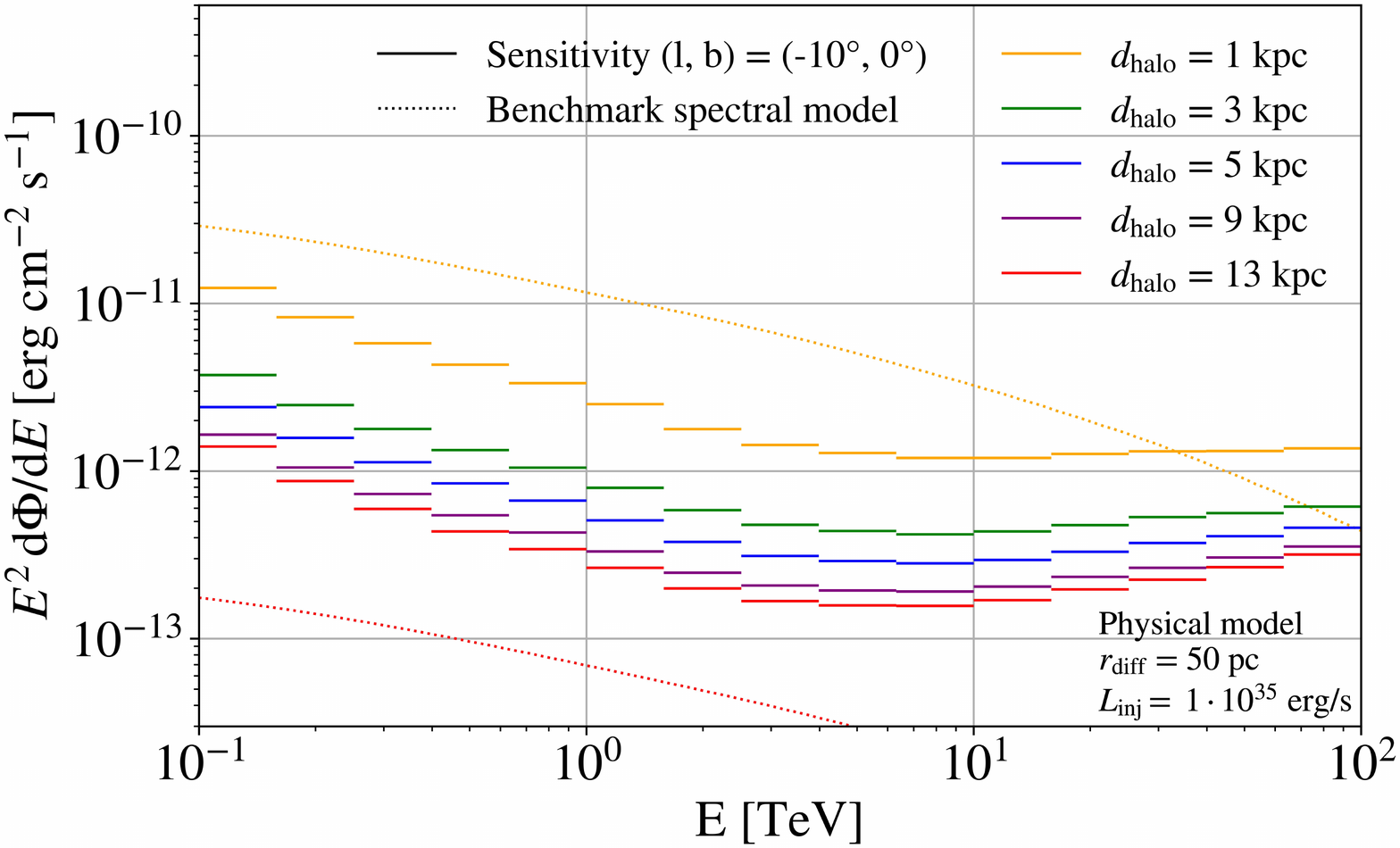}
\par\end{centering}
\caption{Differential spectral sensitivity to extended sources for the Galactic Plane survey observations, focusing on a 6  degree region centered at $(l,~b)=(-10^{\circ}, 0^{\circ})$. \emph{Left}: Differential sensitivity to extended Gaussian sources with extensions ranging between $0.0^{\circ}$ and $1.0^{\circ}$. The full lines show the sensitivity for the Galactic Plane survey, while the dashed lines mark the differential sensitivity to an extended Gaussian source assuming a 5h single pointing observation with the Southern array and using \texttt{prod5} IRFs optimized for 20$^{\circ}$ zenith angle. The same colors correspond to the same source extensions. \emph{Right}: Differential sensitivity to our benchmark pulsar halo model for the Galactic Plane survey observations, positioned at different distances from the observer. Dotted lines show the benchmark spectral model for pulsar halos. The same colors correspond to the same source distance. \label{fig:DiffEnSens}}
\end{figure*}
	
%%%%%%%%%%%%%%%%
\subsection{Model-independent sensitivity to extended sources}\label{sens_extsrc}
%%%%%%%%%%%%%%%%%

We start with a general study of the sensitivity of the GPS to extended (Gaussian) sources, since the physical processes governing the development of pulsar haloes are still poorly known and because such a study could be valuable when assessing the detection prospects for other source classes, for instance PWNe, star-forming regions, or dark matter clumps.
%\red{@Francesca - we could add here some words about what expected DM clump extensions would be.}
% PIMA: So we agreed to not go too deep into that topic right ?

In Fig. \ref{fig:DiffEnSens} (left panel), we show the spectral sensitivity to two-dimensional Gaussian intensity distributions with various energy-independent extensions. The sensitivity degrades with an increasing source size by about an order of magnitude over most of the energy range when going from the point-like case up to $\sigma=1.0^{\circ}$. %We remind that these sensitivities hold for the planned GPS exposure at position $(l,b) = (-10^{\circ}, 0^{\circ})$.

The spectral sensitivity curves presented above do not provide any information about the angular size over which a given extended source is significantly detected. We therefore display in Fig. \ref{fig:ang_decomp_model_comparison1} {(left panel)} curves showing angular sensitivity {to uniform ring emission} in three different energy ranges: $0.1-1$\tev, $1-10$\tev, and $10-100$\tev. For details  on how such sensitivities are calculated see Sec.\ref{data:det}. 

%\pima{Not sure we need the single pointing examples, we could go directly for GPS-like exposure, in different parts of the plane. That's already quite many plots.}
%\fig{3, sensitivity for representative source extensions for GPS in different sky regions (with  and without the IEM, and by including benchmark value for syst uncertainty)}

%%%%%%%%%%%%%%%%%%%%
\subsection{Sensitivity to a physical model of pulsar halos}\label{sens_psrhalo}
%%%%%%%%%%%%%%%%%%%%

%\gabi{Here we focus on a physical model of TeV haloes (see Section XX) and calculate the CTA sensitivity in this, more realistic case. As discussed previously the radial profile of sources in this model is set by the assumed diffusion coefficient and is therefore energy dependent.}

%\pima{Leave effect of halo size out for the moment as it is degenerate with injection index and not sure how to best disentangle that. So let's use 50pc as default, some investigation needed to compare 30-50-80pc cases.}

In this section  we turn instead to simulating pulsar halo emission following  the physical model proposed in \citet{Martin:2022}.  In this case, the gamma-ray extension of the halo is set by the assumed setup of the model and is in the general case energy dependent, in contrast to our Gaussian scenario above.

%%%%%%%%%%%%%%%
\subsubsection{Spectral sensitivity}
%%%%%%%%%%%%%%%%

The spectral sensitivity, shown in the right panel of Fig. \ref{fig:DiffEnSens} for a 50\,pc suppressed diffusion region, degrades with an increased extension of the halo, resulting from a closer location, as also found for the general Gaussian source case (see Sec. \ref{sens_extsrc}). The sensitivities to physical halo models are roughly consistent with those of a Gaussian source (left panel of Fig. \ref{fig:DiffEnSens}) of the corresponding angular size (the containment radius of the halo emission, which is energy-dependent but typically of the order of the suppressed diffusion region extent, placed at the corresponding $d_{halo}$ distance; as a reference, the angular radius of a $50$\,pc extent placed at a distance of $1~(5, ~13)$\,kpc, is $1.43^\circ$ $(0.29^\circ, 0.11^\circ)$). The sensitivities to the physical halo model, however, tend to be flatter with increasing energy as a result of the source size shrinking with energy, owing to stronger energy losses with increasing particle energy, which counterbalances the otherwise degrading sensitivity of the instrument beyond about 10\tev.

%\gabi{We also show the flux levels expected for our 'benchmark' individual halo placed at distances of 1 and 13 kpc. Plot shows that nearby haloes would  be detectable with good energy resolution up to tens of TeV (for benchmark injection power of XX). Refer to  population study section.}

%\pima{Building upon Gabi's text...} 

For increasing distances, the sensitivity initially improves roughly with the inverse of the distance, because the same halo signal is mixed with instrumental background over a smaller area, but that trend tends to flatten for distances $\gtrsim 4-5$\,kpc, when the typical angular size of the halo becomes comparable to or smaller than the angular resolution of the instrument. In the meantime, the flux from a halo decreases with the inverse of the distance squared. 

Figure \ref{fig:DiffEnSens} displays the flux levels of our reference individual halo model, for comparison to the computed spectral sensitivities. The reference halo model features a present-day injection luminosity of $10^{35}$\punit, or about five times the one involved in Geminga \citep{Martin:2022}. We did not investigate prospects for haloes closer than 1\,kpc because their typical extent would exceed the latitude extent of the survey. Such objects would require additional, dedicated observations. Our results suggest that fine spectral studies from a few hundreds of GeV to a few tens of TeV would be possible for nearby haloes at distances $1-3$\,kpc and involving injection powers a few times above that inferred for Geminga, or about $10^{35}$\punit. We will see in Sect. \ref{phy:popu} that two to three dozens of pulsars with properties possibly in that range are known. At larger distances, above 5\,kpc, the power requirement for fine spectral studies increases to about $10^{36}$\punit or more, which certainly reduces the pool of possible targets since not many middle-aged pulsars have retained such a power.

%\pima{Assess the impact of latitude, systematics, IE updated models, and distance to the halos. Compare with typical model spectra in selected cases.}

%\gabi{Which of these figures we want to show:}\fig{Impact of halo distance for default CR+MINIMAL background}\fig{Impact of IE for 2-3 selected distances like 1/3/5kpc}\fig{Impact of latitude for 2-3 selected distances like 1/3/5kpc}\fig{Impact of systematics. Or in same plot, as IEM}

%%%%%%%%%%%%%%%%
\subsubsection{Angular sensitivity}
%%%%%%%%%%%%%%%

%In this section we  examine the capability of the CTA to resolve the angular profile of pulsar halos. Starting from a given halo profile, we calculate the uncertainty with which each ring would be detected, (when normalization of the inner rings to the one in question is marginalized over). Such angular decomposition is presented in Fig. \ref{fig:ang_decomp_model_comparison1} (right) and leads consistent results  with our  model independent approach based on individual uniform rings, shown in the left panel. 
%\gabi{The figures show that the CTA will be able to resolve the angular profile with at least two angular bins, for all chosen energy bins, in case of close by halos. MORE TAKE HOME MESSAGES.}

%\pima{Building upon Gabi's text...} 
In this section, we assess the capability of the CTA to resolve the angular profile of pulsar halos. This is done in two steps: we first compare the intensity distribution of various halo models to estimated model-independent angular sensitivity curves in three energy bands, computed following the methodology exposed in Sect. \ref{data:sens}; we then perform an angular decomposition from mock observations of these given halo models, to confirm the prospects suggested in the first step. 

\begin{figure*}
\begin{centering}
\includegraphics[width=0.48\linewidth]{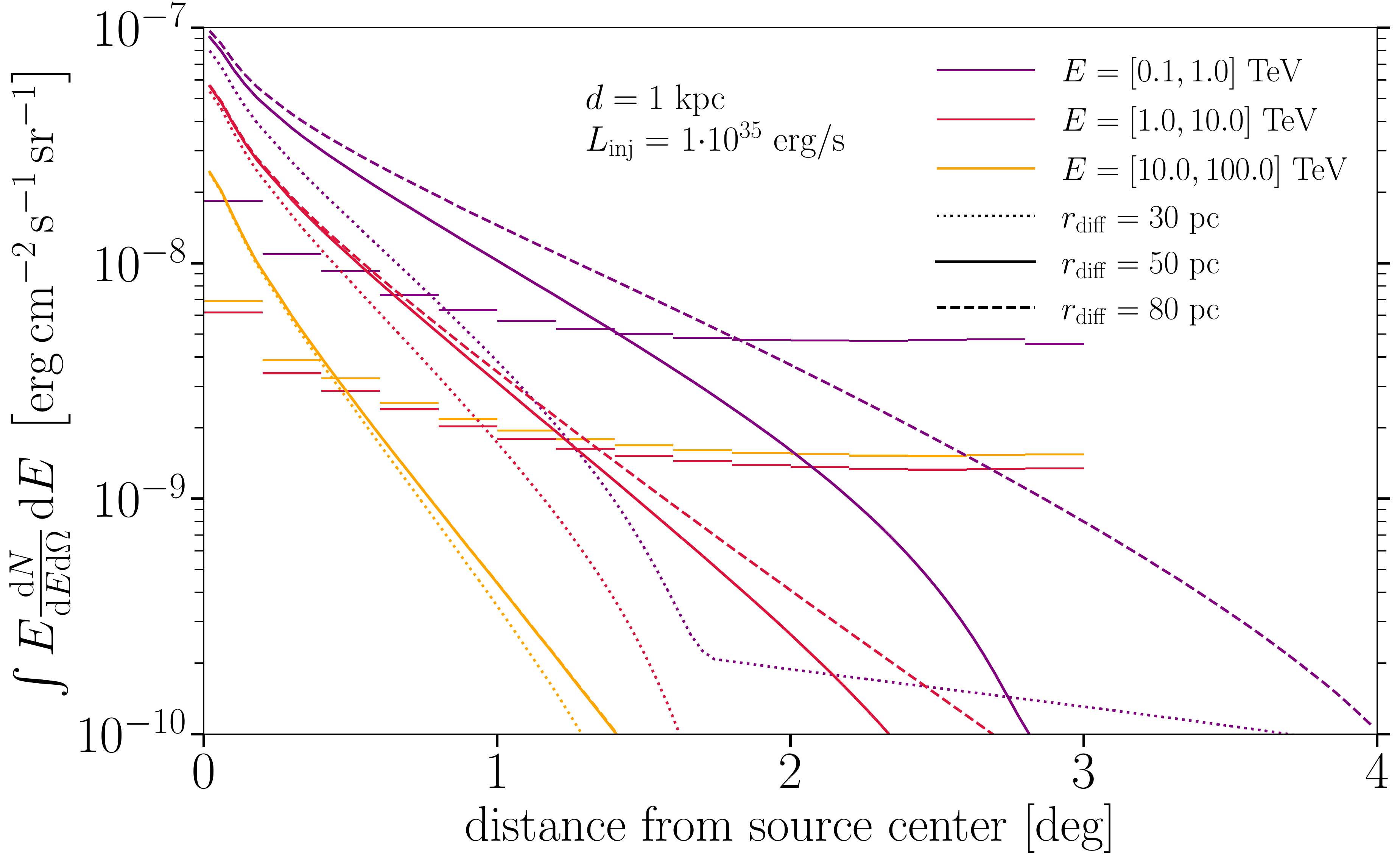}
\includegraphics[width=0.48\linewidth]{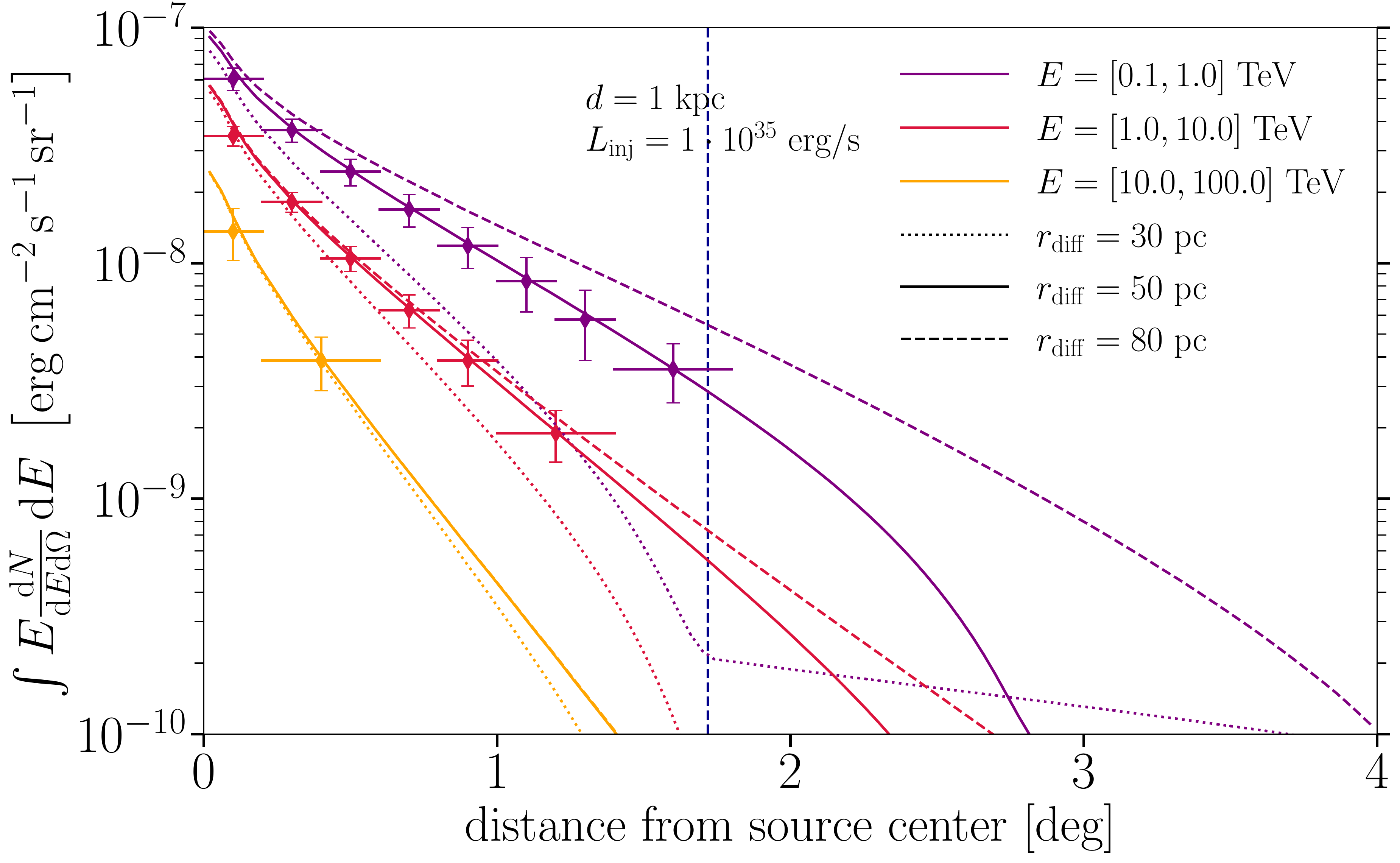}
\includegraphics[width=0.48\linewidth]{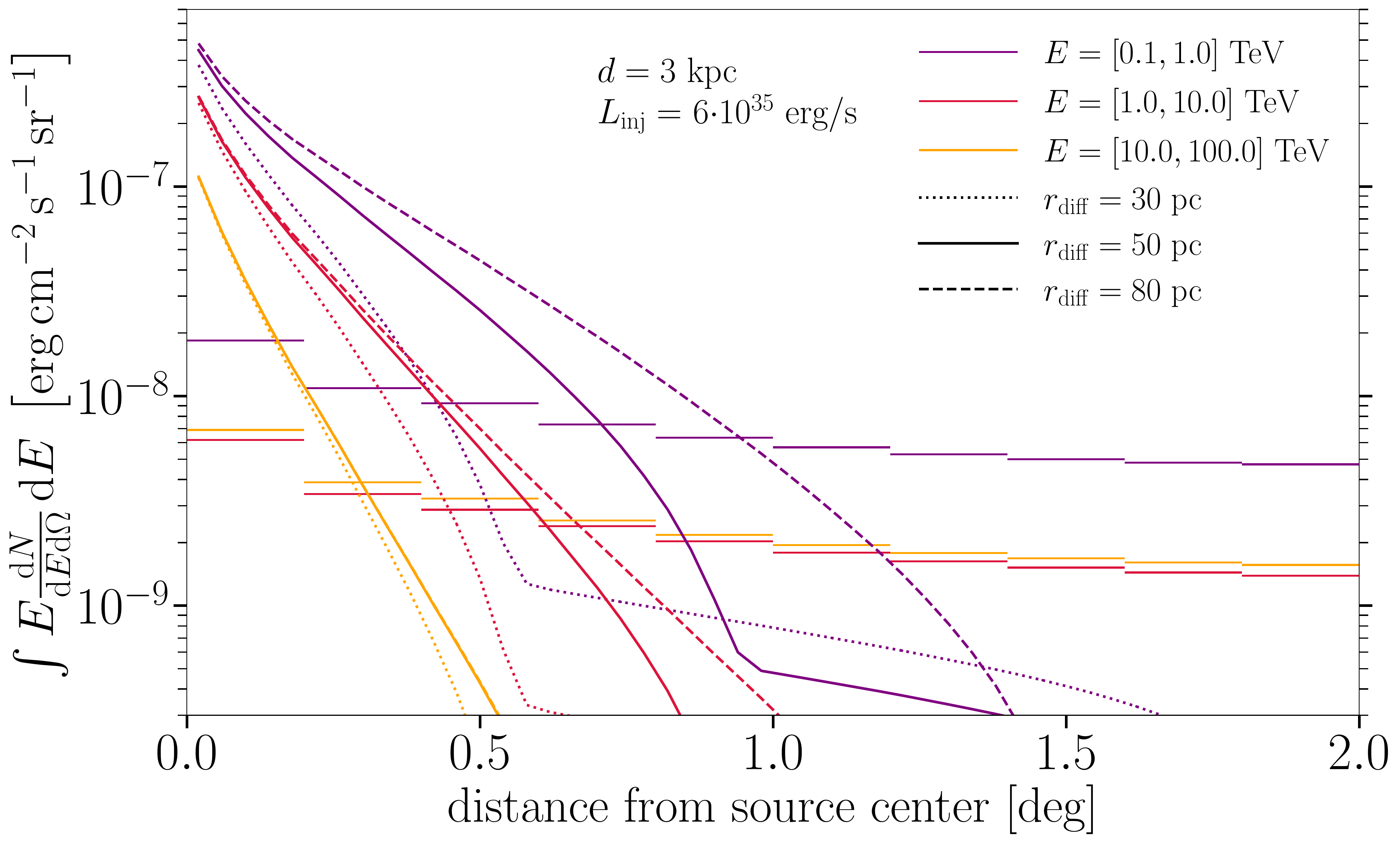}
\includegraphics[width=0.48\linewidth]{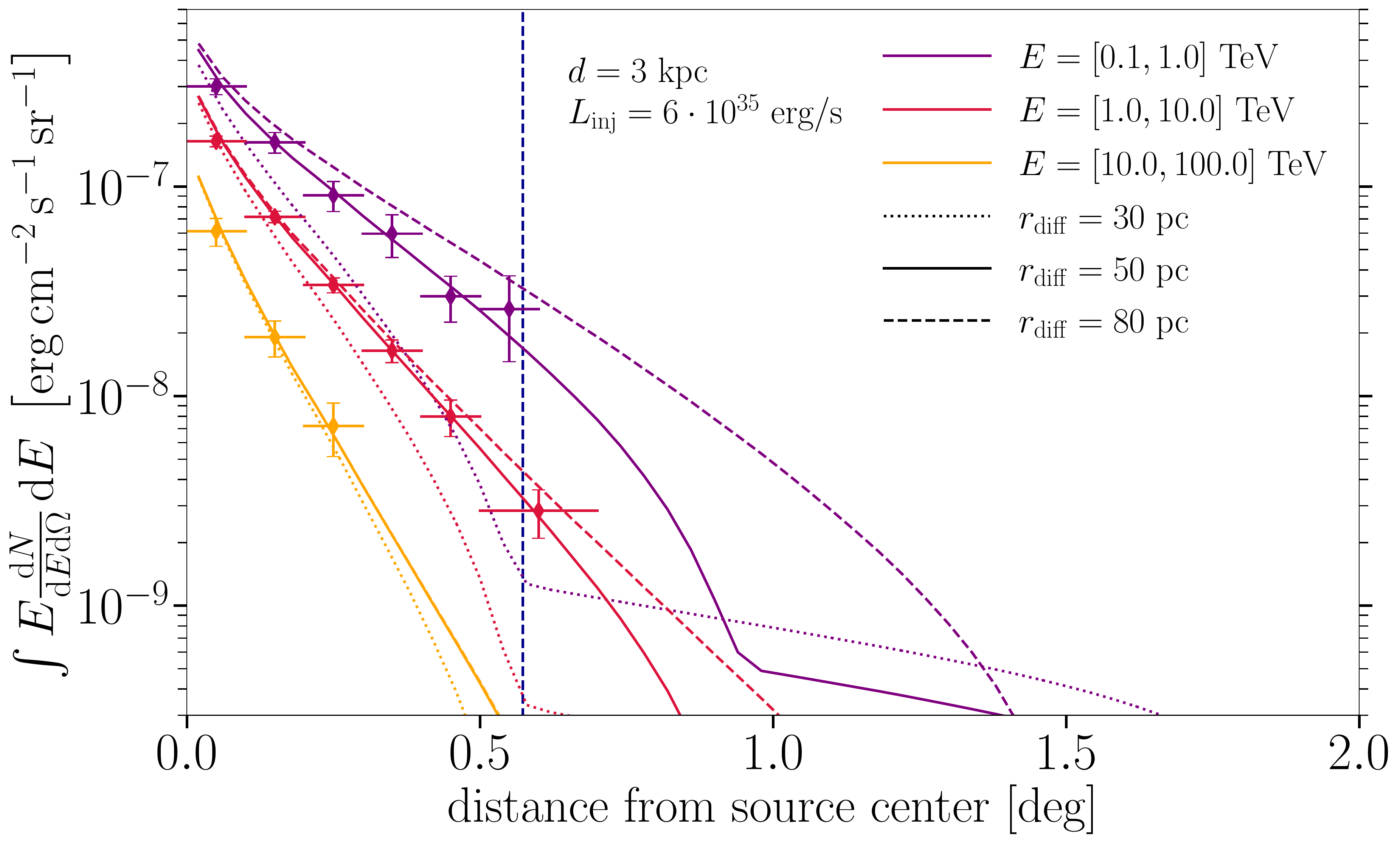}
\par\end{centering}
\caption{Model-independent (\emph{left}) and model-dependent (\emph{right}) angular sensitivity in the three energy bands (0.1 - 1 TeV, 1 - 10 TeV and 10 - 100 TeV), plotted together with the predicted intensity of pulsar halo models with a diffusion region size of 30 pc (dotted lines), 50 pc (full lines) and 80 pc (dashed lines) located at $(\ell, b) = (-10^{\circ}, 0^{\circ})$ at 1 kpc distance from the observer and an injection power of $L_{\mathrm{inj}} = 1\cdot10^{35}$ erg/s (\emph{top}) or at 3 kpc distance from the observer and an injection power of $L_{\mathrm{inj}} = 6\cdot10^{35}$ erg/s (\emph{bottom}) studied under the conditions of CTA's Galactic plane survey. The injection power has been chosen so that the haloes are above the yellow line in Fig.~\ref{fig:halos_powers_sensitivity} allowing for an angular decomposition out to at least 30 pc (denoted by a blue vertical line). \emph{Left}: Angular sensitivity to uniform brightness annuli with a fixed width of $0.2^{\circ}$. \emph{Right}: Prospects for an angular decomposition study of the benchmark pulsar halo with a diffusion zone size of 50 pc.
%located at $(\ell, b) = (-10^{\circ}, 0^{\circ})$ at a distance of 1 kpc (\emph{top}) or 3 kpc (\emph{bottom}) and an injection power of $\dot{E} = 1\cdot10^{35}$ erg/s (\emph{top}) or $\dot{E} = 6\cdot10^{35}$ erg/s (\emph{bottom}) studied under the conditions of CTA's Galactic plane survey. The injection power has been chosen so that the haloes are above the yellow line in Fig.~\ref{fig:halos_powers_sensitivity} allowing for an angular decomposition to at least 30 pc. 
The halo has been decomposed into annuli following the prescription in Sec.~\ref{data:sens} with a minimal width of $0.2^{\circ}$ for $d = 1$ kpc and $0.1^{\circ}$ for $d = 3$ kpc as indicated by the horizontal error bars. The vertical error bars denote the statistical uncertainty of the reconstructed flux within the found annulus.
%For comparison, we show the predicted intensity of halo models with a diffusion size of 30 pc (dotted lines) and  80 pc (dashed lines) at the same distance from the observer. 
%The vertical yellow line indicates the outer boundary of a disc of 30 pc radius for the respective halo in degree.
\label{fig:ang_decomp_model_comparison1}
%\pima{The angular decomposition for the 3kpc case could be done with a much finer angular bining of 0.1 instead of 0.2 deg. And the last point can probably be dropped.}
}
\end{figure*}

The left panels in Fig. \ref{fig:ang_decomp_model_comparison1} show angular sensitivity curves obtained for a set of concentric uniform-brightness annuli. These are compared to our reference halo model at 1 and 3\,kpc, for different suppressed diffusion region sizes, and scaled for injection powers of $10^{35}$\punit and $6 \times 10^{35}$\punit, respectively. These normalizations were chosen so that a meaningful angular decomposition over a sufficient number of angular bins could be achieved. In the 1\,kpc case, the halo profile for a 50\,pc suppressed diffusion region lies above sensitivity out to about $1.5\deg$ from the center in the two lower energy ranges, and only out to $0.5\deg$ in the higher energy range. This corresponds to lengths of 26 and 9\,pc at a distance of 1\,kpc, respectively. These statements are dependent on the extent of the suppressed diffusion region, especially at the lowest energies. Overall, the CTA GPS should allow us to probe the $0.1-10$\tev intensity distribution of a pulsar halo with power $10^{35}$\punit and at distance 1\,kpc over an extent comparable to that reached by HAWC for Geminga \citep{Abeysekara:2017b}. This is confirmed by the actual angular decomposition performed from mock observations based on the same halo model (see the top right panel in Fig. \ref{fig:ang_decomp_model_comparison1}). In the 3\,kpc case, qualitatively similar results are obtained, at the expense of a six times higher injection power.

%\gabi{Elena Amato suggested to explore what is the max number of energy bins for which we could tell the angular distribution. Do we place that  in Appendix? How do we approach  it?}

%\gabi{Do we want to show a plot as a function of observation time, for dedicated surveys} \pima{Why not, but let's see first what GPS prospects are}

%\pima{A possible first plot here is that of of angular sensitivity in 0.1-0.2deg uniform-brightness concentric rings, compared to a typical model profiles for haloes that would be detectable and sufficiently luminous to be decomposed out to a significant physical distance like 20-30pc, but we need to check first that results are consistent.}

%\fig{Plot of minimum luminosity to achieve angular decomposition in 5-50TeV band out to at least 30pc, as function of distance to halo. Results should be nearly the same for 30/50/80pc sizes, to be confirmed, and if so we move forward with 50pc as default size. Investigate here the effect of changing band, from 5-50TeV like HAWC to 0.1-1TeV which would be the added value from CTA !}

%\fig{Then, for a set of selected distances like 1 and 3kpc, we can plot the angular decomposition of a sufficiently luminous halo, and overlay model profiles, e.g. for different diffusion coefficients or sizes, to illustrate the constraining power.}

%%%%%%%%%%%%%%%%
\subsubsection{Robustness of results} \label{sec:robustness}
%%%%%%%%%%%%%%%

\begin{figure*}
\begin{centering}
\includegraphics[width=0.48\linewidth]{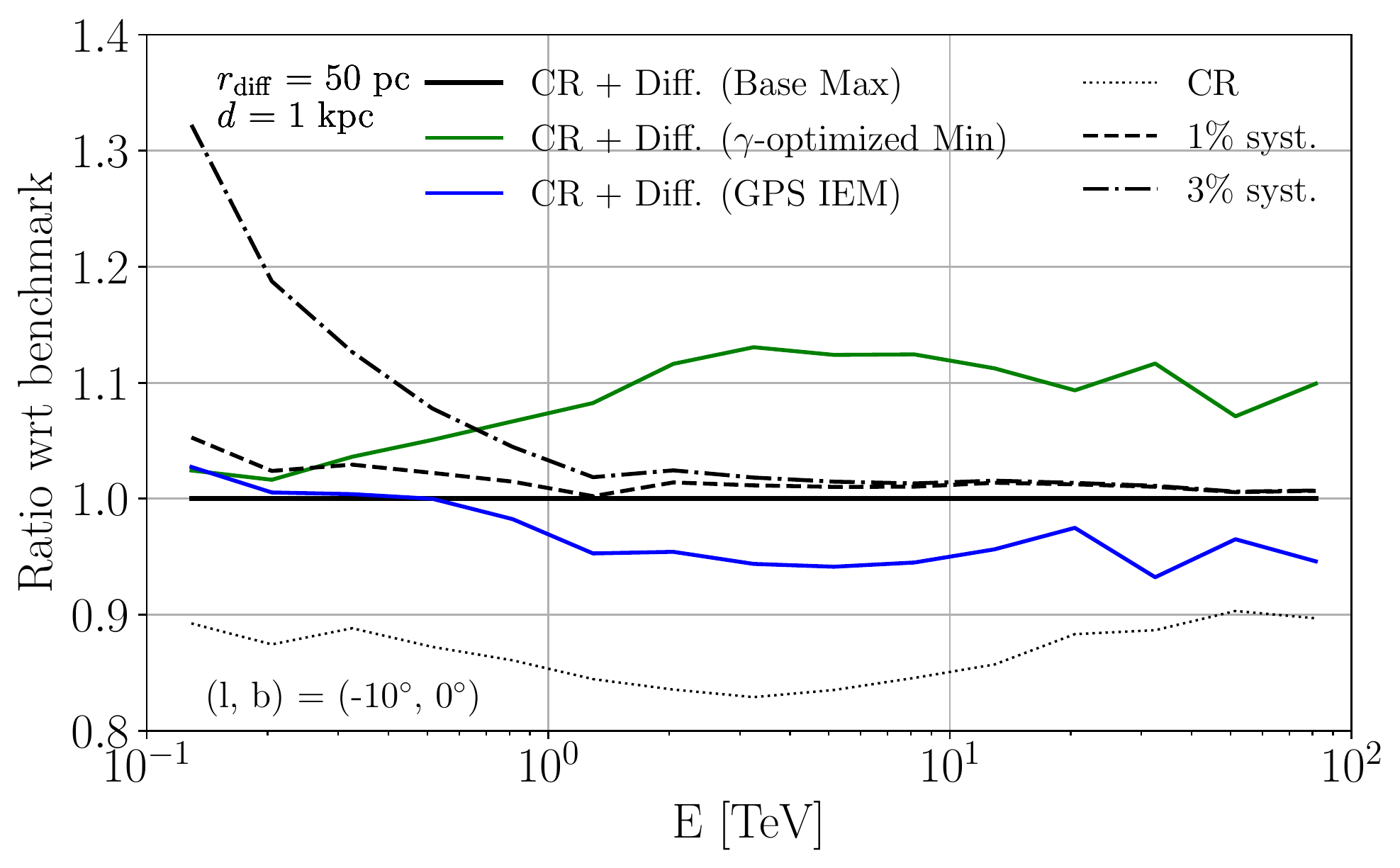}
\includegraphics[width=0.48\linewidth]{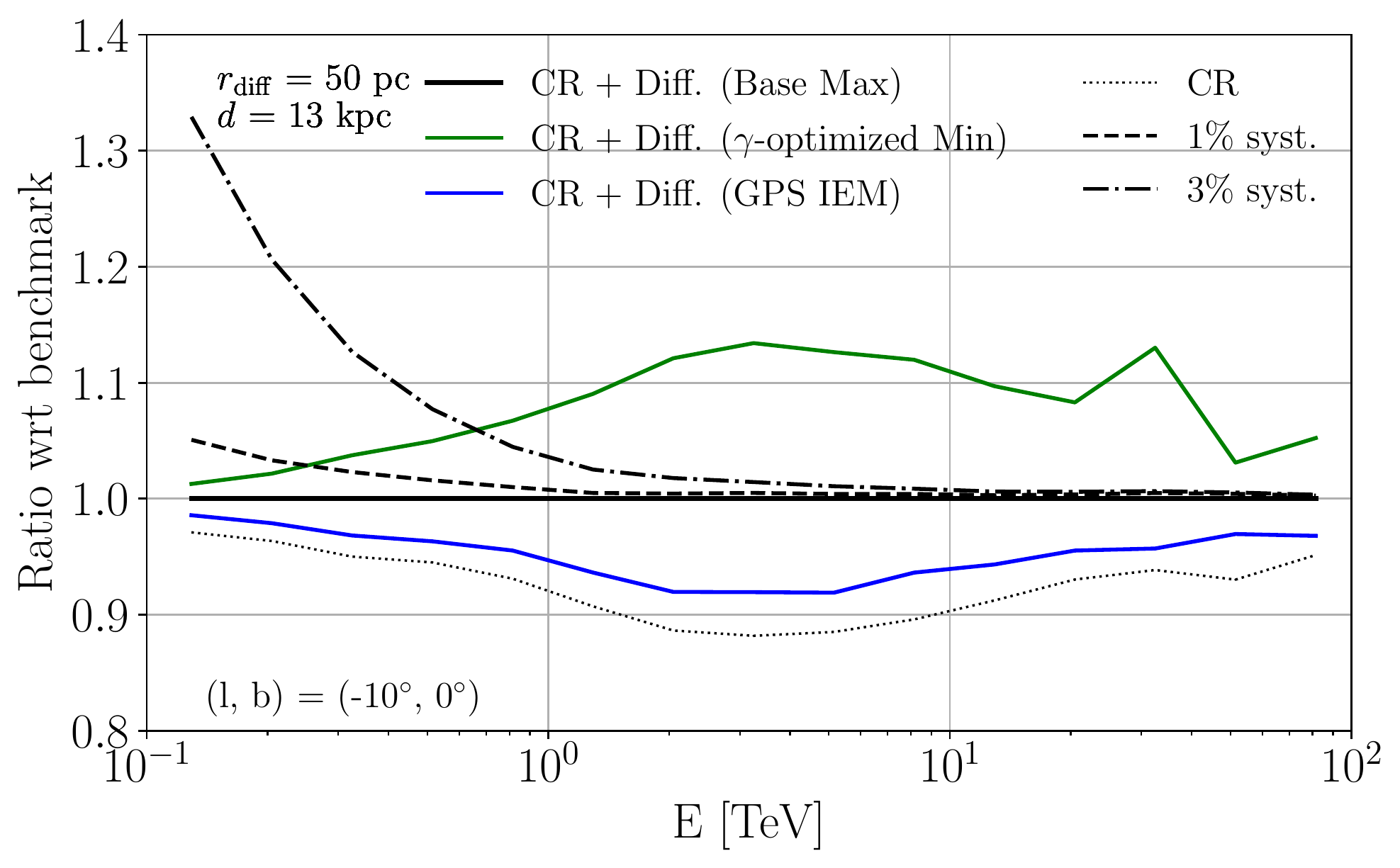}
\par\end{centering}
\caption{The ratios of the differential energy sensitivity with respect to the benchmark model for a halo at 1 kpc (left) and 13 kpc (right), considering different IEMs (green and blue line) or no added IE (black thin dotted line). The effect of added systematics on the benchmark model is shown at 1\% and 3\% levels in black dashed and black dash-dotted lines, respectively.} 
\label{fig:robustness}
\end{figure*}

In this section we explore how dependent our results are on i) the assumption we make on the IE model and ii) instrumental systematic uncertainties. We address this issue by quantifying how our benchmark  sensitivity to haloes changes when relaxing particular assumptions. Specifically, we focus on two cases: a significantly extended halo positioned at 1 kpc (Fig. \ref{fig:robustness}, left), and a point-like halo positioned at 13 kpc distance (Fig. \ref{fig:robustness}, right).   

{\bf Impact of the IE model:} In  Fig. \ref{fig:robustness} we explore the impact of three different assumptions: no IE case (where only the CR background is present), $\gamma$-optimized Min model from \cite{Luque:2022buq} and the IE model that is used in GPS publication (see \cite{Remy:2022}). We observe that, as expected, the impact of IE is larger in the case of an extended source (left panel) but it is limited to less than $20\%$ impact. Note that here we assume that we know the true IE model. In a more realistic case where data are produced with one model (say Base Max), and modeled with another (say $\gamma$-optimized Min), the sensitivities can degrade significantly. In Appendix \ref{app:IEMfit}, we explore the worsening of sensitivities from using different combinations of IE models for data production and modeling. We find that the sensitivities degrade by a factor of up to to 4.

{\bf Systematic uncertainty:} The template likelihood analysis adopted here critically depends on having appropriate models for the emission components. However, many instrumental effects, especially those that affect the observations within the field of view on smaller scales are notoriously hard to model, but could have significant  impact on  the analysis (see discussion in \cite{CTA:2020qlo}). While the exact level or scale of such potential effects are unknown before the construction of the instruments, we follow \cite{CTA:2020qlo} in choosing the scale of 0.1$^{\circ}$ (close to the PSF size) and plot its impact for the magnitude of $1\%$ (which is the target goal of the CTA) and $3\%$. We note that, as expected the impact of  these uncertainties is the highest at low energies (since higher energies are statistics dominated) and are limited to $\leq 10\%$ at around 1 TeV where we  have the best sensitivity to our sources.

%\gabi{Shall we discuss impact of longitude/latitude position  of a halo?}

%%%%%%%%%%%%%%
\section{Physics prospects}\label{phy}
%%%%%%%%%%%%%%%

%%%%%%%%%%%%%%%%
\subsection{Insights for individual halos}\label{phy:indiv}
%%%%%%%%%%%%%%%%%%

When using the halo around Geminga as a canonical instance of the phenomenon (especially in terms of diffusion suppression level), most of the signal above a few TeV is contained within 20-30\,pc of the pulsar and is thus little sensitive to the full extent of the suppressed diffusion region, whether it be 30, 50, or 80pc. This is illustrated in Fig. \ref{fig:ang_decomp_model_comparison1} by the set of yellow model curves, and to a lesser extent by the set of red model curves. These plots show that, even for relatively powerful pulsars with injection powers in the $10^{35-36}$\punit range and located at a few kpc from us, CTA GPS observations are hardly sufficient to discriminate between different extents of the haloes in the core $1-10$\tev range, and definitely too shallow to do so at higher $10-100$\tev energies. At these energies, however, the CTA GPS should enable fine spectral studies of haloes with these properties, reaching above $20-30$\tev for the most powerful and/or closest objects. This will be useful to constrain key parameters of the phenomenon like the injection spectrum or the momentum dependence of suppressed diffusion.

Conversely, there is more potential in the $0.1-1$\tev band to constrain the halo size, because particles radiating in this band have a larger propagation range and fill the halo out to larger distances, to a point that the corresponding emission morphology is sensitive to the location of the suppressed diffusion region boundary. This is illustrated in the right panels of Fig. \ref{fig:ang_decomp_model_comparison1} by the set of purple model curves, to be compared with the predicted angular decomposition displayed as purple dots with error bars. This is complemented by the prospect of fine spectral studies down to or even below 100\gev, as illustrated in Fig. \ref{fig:DiffEnSens}.

When it comes to the study of pulsar halos, CTA can therefore be expected to nicely complement HAWC and LHAASO by, among other things, offering an extension of the energy coverage below 1\tev, in a regime where the emitting particles are proportionally less affected by energy losses and can thus probe the entirety of the suppressed diffusion region. The longer lifetime of these particles also implies that the pulsar's proper motion should start to have an impact on the morphology of the halo. This could be an additional advantage as it may provide a specific spectro-morphological signature, or a disadvantage as spilling the emission over a larger patch of the sky may reduce the brightness of the signal and increase source confusion. These statements remain qualitatively valid for lower levels of diffusion suppression, such as those possibly involved in Monogem. This would only shift the energy below which the survey is sensitive to the diffusion region extent to higher values, which makes the case even more interesting as discernible effects would be accessible at core energies for CTA where the sensitivity is the highest.

%%%%%%%%
\subsection{Accessible fraction of a Galactic population}\label{phy:popu}
%%%%%%%%

\begin{figure*}
\begin{centering}
\includegraphics[width=0.70\linewidth]{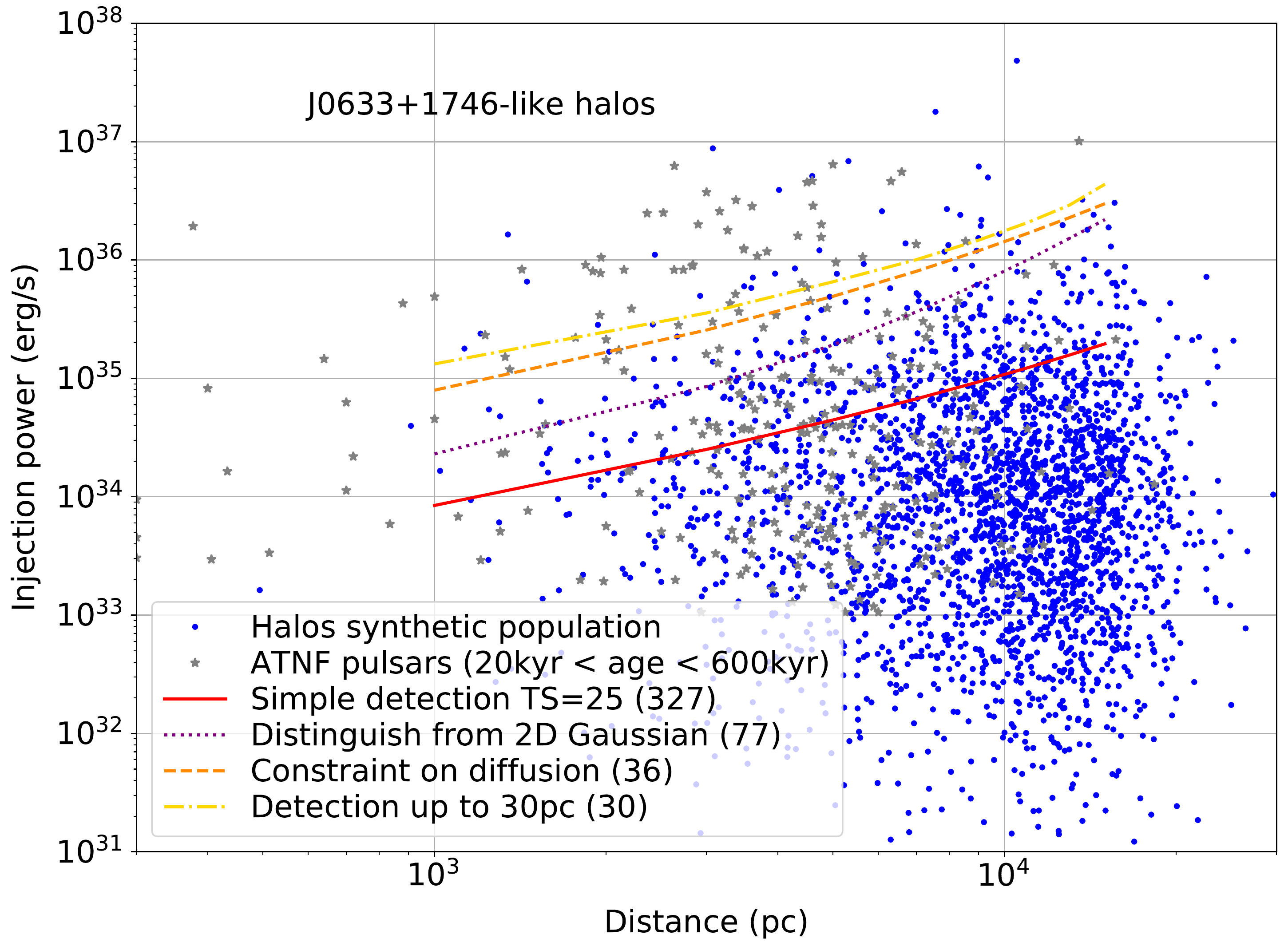}
\includegraphics[width=0.70\linewidth]{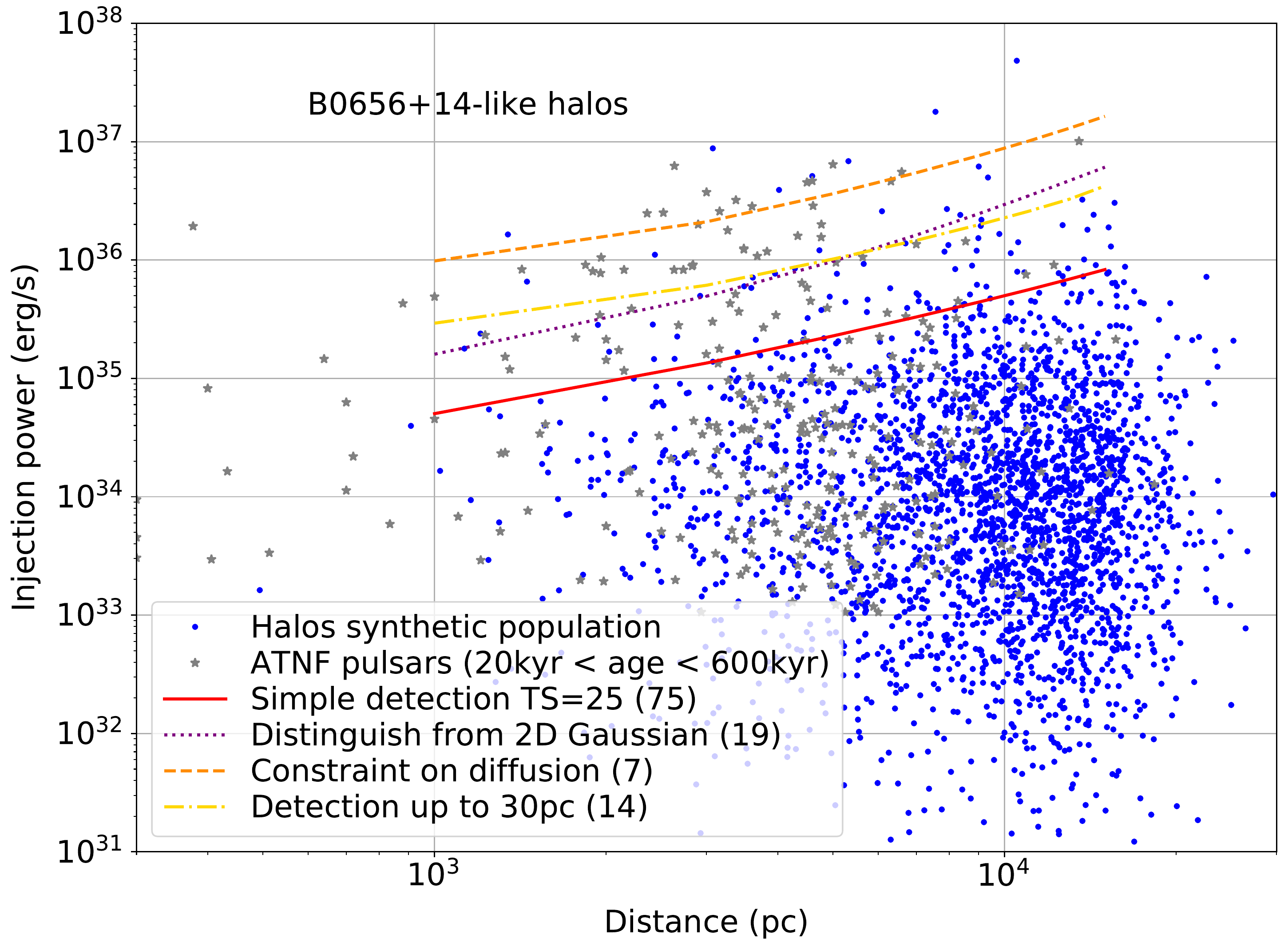}
\par\end{centering}
\caption{Sensitivity of the survey in terms of particle injection power as a function of distance, for our two representative halo models: with suppressed diffusion typical of J0633+1746 (\textit{top}) or of B0656+14 (\textit{bottom}). The blue points marks the locations of mock haloes from our population synthesis. The gray stars mark the locations of pulsars listed in the ATNF data base and having characteristic ages in the 20-600\,kyr range. The curves represent sensitivities to various source features: in solid red line, overall detection over the full 0.1-100\,TeV range with TS=25; in purple dotted line, significant separation of the true halo model from a 2D Gaussian intensity distribution; in yellow dot-dashed line, significant detection of emission beyond 30\,pc from the pulsar; in orange dashed line, significant separation of the true halo profile from that obtained with a 50\% higher diffusion coefficient. For each sensitivity curve, the numbers in parentheses indicate the number of mock haloes lying above the curve.} 
\label{fig:halos_powers_sensitivity}
\end{figure*}

Based on the analyses presented above of the sensitivity of the survey to representative haloes at various distances, we estimate the fraction of the whole halo population that should be within reach of CTA and, going beyond mere detection, that would allow deeper investigations of the objects.

For each representative halo model (J0633+1746-like or B0656+14-like), and for each distance in the 1-15\,kpc range, we performed a series of analyses to compute the following quantities:
\begin{enumerate}
    \item Injection power such that the simulated halo signal is detected with a TS of 25 over the full energy range, using the true halo model in the fit process.
    \item Injection power such that a fit of the simulated halo signal with the true halo model is significantly better than a fit with a simple energy-independent 2D Gaussian intensity distribution.
    \item Injection power such that a fit of the simulated halo signal with the true halo model is significantly better than a fit with the true model clipped to zero beyond a distance of 30\,pc from the pulsar.
    \item Injection power such that a fit of the simulated halo signal with the true halo model is significantly better than a fit with an alternative halo model having a 50\% higher suppressed diffusion coefficient.
\end{enumerate}
In the last three cases, significant means twice the difference in log-likelihood of 16 (which corresponds to a $4\sigma$ effect when a variation in only one parameter in nested models is tested, which is not strictly the case in all situations listed above). The 30\,pc limit used as criterion in the third test is the typical distance up to which the HAWC observations made it possible to determine a radial intensity profile for J0633+1746 and  B0656+14 in Ref. \cite{Abeysekara:2017b}. These analyses were done in a simplified framework where observation simulations are based on a source model for the ROI consisting in the halo model and the instrumental background only, while model-fitting to the mock data in the test hypothesis is based on the true models for the halo energy-dependent intensity distribution and instrumental background. In actual data analysis conditions, the source properties are not known a priori, there is source confusion from other unknown astrophysical signals in the field, and instrumental background is not fully under control. So the final results presented below in terms of fraction of the accessible population should be considered as optimistic upper limits.

In Fig. \ref{fig:halos_powers_sensitivity}, we displayed these requirements in terms of injection power as a function of distance and compare them to the distribution of mock haloes from the population synthesis presented in Ref. \cite{Martin:2022b}. The number of mock haloes accessible to the survey for a given scientific goal (simple detection, detection of emission up to 30\,pc,...) can be evaluated by comparing their actual injection power to that required for a halo at this distance, after correcting for the longitude-dependent sensitivity of the survey. The latter correction is implemented by rescaling our sensitivities, computed for a reference position in the inner Galaxy where the survey sensitivity is at its highest, by the ratio of the targeted survey sensitivity at the halo position to that at our reference position, using the survey performance description in Ref. \cite{Acharya:2019}.

For objects at a distance in the 1-15\,kpc range, about 350 (respectively 100) haloes would be detectable if they are assumed to be J0633+1746-like (respectively B0656+14-like). The plot however illustrates that a much lower number of objects are expected to allow deeper investigation of their physical properties. Only 30 (respectively 14) J0633+1746-like (respectively B0656+14-like) haloes can be significantly detected up to 30\,pc, which should make it possible to perform meaningful angular decomposition of their emission, as illustrated in Fig. \ref{fig:ang_decomp_model_comparison1}. Similarly, significant constraint on the suppressed diffusion coefficient can be achieved for only 40 haloes in the J0633+1746-like model setup, and four times fewer objects with the B0656+14-like one.

The prospects for mere detection of J0633+1746-like pulsar haloes are about twice those obtained in Ref. \cite{Martin:2022b} from a simple flux criterion (about 160 objects, see their Table 2). This is not surprising as the present estimates are based on a method that exploits the full spectro-morphological signature of haloes over a broad energy range. The number of detectable J0633+1746-like pulsar haloes thus becomes comparable to the number of detectable PWNe presented in Ref. \cite{Remy:2022}.

%%%%%%
%\subsection{Source confusion}\label{xx}
%%%%%

%\gabi{Shall we comment explicitly? Maybe analyse a sky region with two comparable overlapping halos? Or make a pie-chart as the one shown at the CTAC? Or just remove this section and comment in Sec. \ref{phy:diff} or  conclusions that confusion is not taken into account?}

%%%%%%
\subsection{Promising pulsar candidates}\label{phy:cand}
%%%%%

In the plots of Fig. \ref{fig:halos_powers_sensitivity}, we displayed the positions of known pulsars from the Australia Telescope National Facility (ATNF) data base with characteristic ages in the 20-600\,kyr range (younger pulsars are expected to be in their PWN stage, and older ones to be too faint for detection). Our predicted sensitivity limits can be used to select those ATNF pulsars that could be detectable if they were to develop a halo, under the assumption that 100\% of their spin-down power is injected into nonthermal particles. 

We present in Table \ref{tab:atnf-cand-geminga} a list of those 122 pulsars that should be detectable according to our sensitivity estimate, and under the assumption that they develop a Geminga-like halo. We specify for each of them whether they should allow deeper investigations beyond simple detection. Out of 122, 65 can be distinguished from a two-dimensional energy-independent Gaussian intensity distribution and 41 can be detected up to 30\,pc at least. These numbers are comparable to those obtained from the population synthesis. Not all pulsars listed in this table are expected to be good halo candidates, in particular because some may still be in their PWN stage and have unambiguously been identified as such already, or have actual injection efficiencies lower than 100\%, but the sample is a good starting point for a selection of targets.

\subsection{Diffuse emission from the unresolved population}\label{phy:diff}

\begin{figure*}
\begin{centering}
\includegraphics[width=0.70\linewidth]{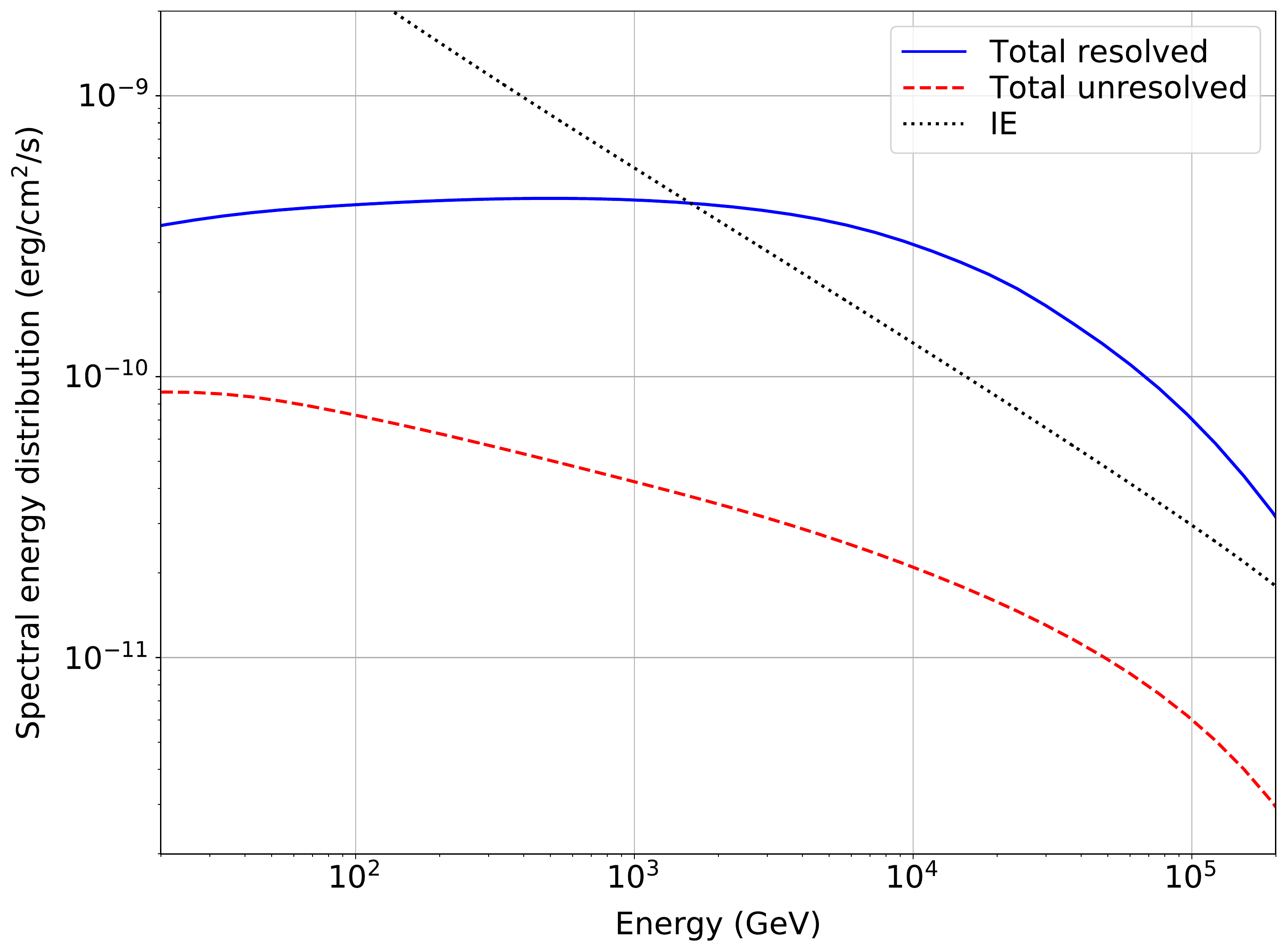}
\includegraphics[width=0.70\linewidth]{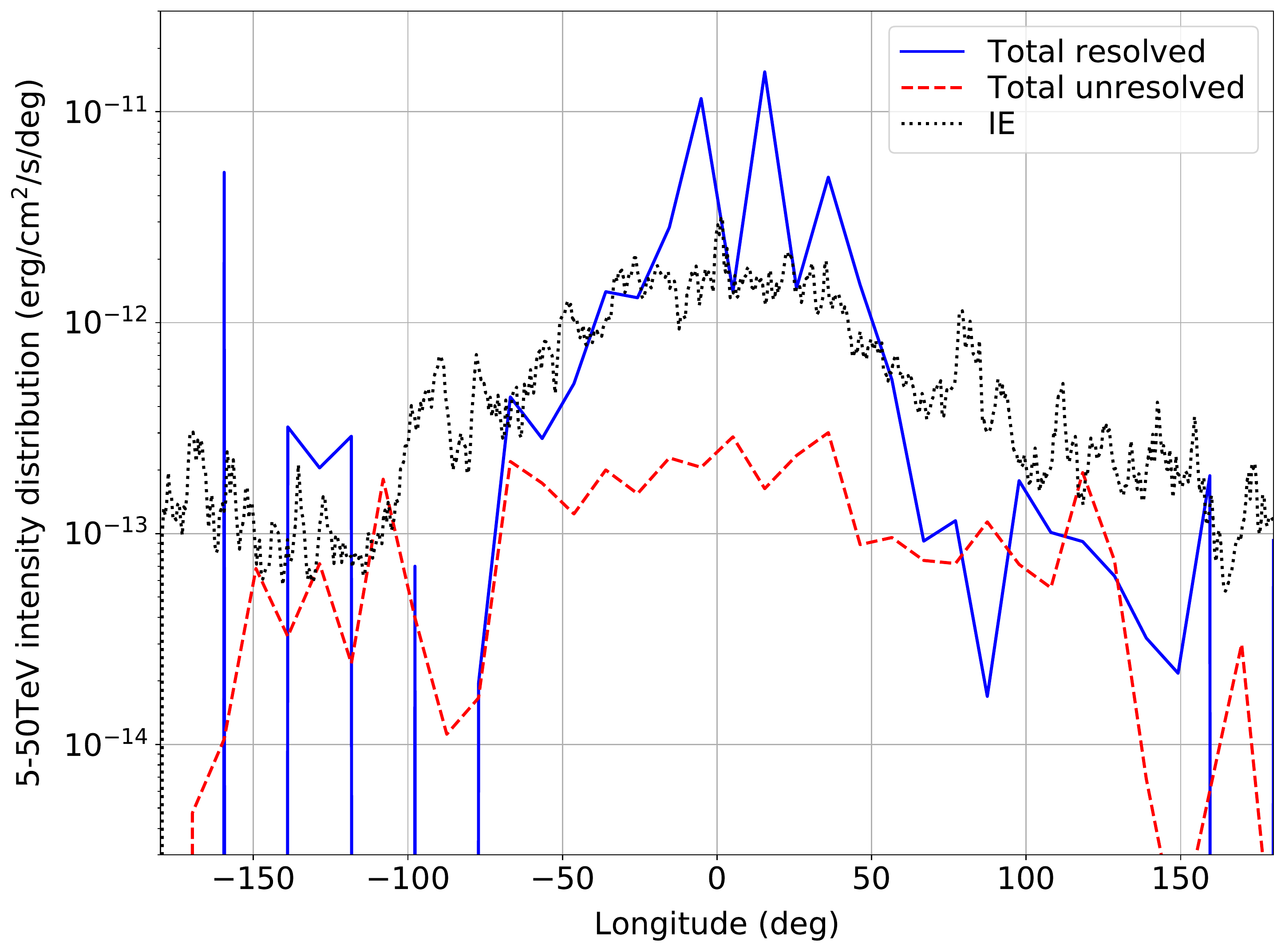}
\par\end{centering}
\caption{Spectrum and longitude distribution of the total emission from resolved or unresolved halos, when the representative halo model features suppressed diffusion typical of J0633+1746. Both are integrated over the survey footprint. For comparison, the total emission from interstellar radiation is also displayed, using the `Base Max' model setup from Ref.~\protect\cite{Luque:2022buq}. The longitude distributions for halo emission are computed over 10-degree bins to smooth out the fluctuations from the actual realization of the population synthesis. 
\label{fig:halos_unresolved_population}}
\end{figure*}

Figure \ref{fig:halos_unresolved_population} shows the properties of the diffuse emission contributed to by the $\sim2000$ or more pulsar haloes that are not detectable individually, for one single realization of the population synthesis and when the representative halo model features suppressed diffusion typical of J0633+1746. 

Integrated over the whole survey footprint, this emission is subdominant compared to that from interstellar radiation, described with the `Base Max' model from Ref. \cite{Luque:2022buq}, with a maximum contribution at the level of $10-20$\% in the $10-100$\tev range. Yet, the longitude distribution shows a more contrasted picture, with regions of comparable $5-50$\tev intensity such as the Cygnus-Cassiopeia portion of the plane, while interstellar emission in the inner Galaxy is almost an order of magnitude above that from unresolved halos.

When haloes are modeled with shallower suppressed diffusion typical of B0656+14, they are on average fainter and more extended, hence less likely to be detected (four times fewer objects are detectable in the survey; see Sect. \ref{phy:popu}). Both effects combine to eventually yield an emission from unresolved haloes that is more intense than when haloes are modeled with suppressed diffusion typical of J0633+1746, and comparable to interstellar emission along the plane.

Last, we emphasize that the above statements hold under the strong assumption that all middle aged pulsars develop halos. If only a small fraction ($5-10$\% of them) do so, as suggested in Ref. \cite{Martin:2022}, haloes become negligible in terms of contribution to the Galactic diffuse emission.

\section{Conclusions}\label{conclusion}

We presented a quantitative assessment of the prospects for the detection and study of the emerging class of pulsar haloes in the observations planned for the CTA GPS. Using a simple phenomenological two-zone diffusion model for individual pulsar haloes and their population in the Milky Way, we simulate a realistic study of these objects in the framework of a full spatial-spectral likelihood analysis of simulated survey observations.

For a halo model setup consistent with the HAWC observations of the halo around PSR~J0633+1746, and under the assumption that all middle-aged pulsars that exited their original nebula develop a halo, we show that about three hundreds objects could give rise to detectable emission components in the survey. Yet, only a third of them could be identified through their energy-dependent morphology, and only one tenth of them would allow the derivation of strong constraints on key physical parameters like the magnitude or extent of suppressed diffusion around the pulsar. These numbers are roughly divided by four when using a model setup consistent with the HAWC observations of PSR~B0656+14 instead.

For pulsar haloes sustained by particle injection power in the range $10^{35-36}$\punit  and located out to a few kpc from us, the CTA GPS observations should enable fine spectral studies from hundreds of GeV or below up to a few tens of TeV. The $0.1-1$\tev band accessible to CTA holds a lot of potential for constraining the transport properties in halos. CTA can be expected to complement HAWC and LHAASO by extending the energy coverage below 1\tev, in a regime where the emitting particles are proportionally less affected by energy losses and can thus probe a larger volume around the pulsar.

Last, we provide a list of known pulsars that could be hosting a detectable (Geminga-like) halo in the GPS, together with information about the likelihood to achieve different analysis goals if it is the case (from simple detection to the derivation of meaningful constraint of the suppressed diffusion characteristics).

\medskip

\section*{Acknowledgements}
\addcontentsline{toc}{section}{Acknowledgements}
We warmly thank Rub\'en L\'opez-Coto and Daniela Hadasch for carefully reading our manuscript and helpful comments.
The work of C.E.~is supported by the ``Agence Nationale de la Recherche'' through grant ANR-19-CE31-0005-01 (PI: F.~Calore). The work of C.E.~has been supported by the EOSC
Future project which is co-funded by the European Union
Horizon Programme call INFRAEOSC-03-2020, Grant
Agreement 101017536.
The work of P.M. is supported by the ``Agence Nationale de la Recherche'' through grant ANR-19-CE31-0014 (GAMALO project, PI: P.~Martin).

\emph{Software:} \textsc{astropy} \citep{2013A&A...558A..33A, 2018AJ....156..123A,TheAstropyCollaboration2022}, \textsc{healpy} \citep{2005ApJ...622..759G, Zonca2019}, \textsc{iminuit} \citep{iminuit}, \textsc{jupyter} \citep{soton403913}, \textsc{matplotlib} \citep{4160265}, \textsc{numpy} \citep{2020Natur.585..357H}, \textsc{scipy} \citep{2020NatMe..17..261V}
%%%%%%%%%%%%%%%%%%%%%%%%%%%%%%%%%%%%%%%%%%%%%%%%%%
\section*{Data Availability}
% Entry for the table of contents, for this guide only
\addcontentsline{toc}{section}{Data Availability} 
The data that support the findings of this study are available from the corresponding authors [C.E., V.V., P.M.] upon reasonable request.

\bibliographystyle{mnras}
\bibliography{biblio_psrhalos,biblio_psrhalos_pierrick,software_bib}

\clearpage
\onecolumn
\begin{longtable}{| c | c | c | c | c |}
\hline
Pulsar & Age (kyr) & Distance (pc) & Spin-down power (erg/s) & Flags \\
\hline
J0002+6216 & 306.0 & 6357 & 1.53$\times 10^{35}$ & D \\
B0114+58 & 275.0 & 1768 & 2.21$\times 10^{35}$ & DMC \\
J0248+6021 & 62.4 & 2000 & 2.13$\times 10^{35}$ & DM \\
B0355+54 & 564.0 & 1000 & 4.54$\times 10^{34}$ & DM \\
B0540+23 & 253.0 & 1565 & 4.09$\times 10^{34}$ & D \\
J0631+0646 & 486.0 & 4583 & 1.04$\times 10^{35}$ & D \\
J0631+1036 & 43.6 & 2105 & 1.73$\times 10^{35}$ & DM \\
J0633+0632 & 59.2 & 1355 & 1.19$\times 10^{35}$ & DM \\
J0729-1448 & 35.2 & 2679 & 2.81$\times 10^{35}$ & DM \\
B0740-28 & 157.0 & 2000 & 1.43$\times 10^{35}$ & DM \\
J0834-4159 & 448.0 & 5508 & 9.51$\times 10^{34}$ & D \\
J0855-4644 & 141.0 & 5638 & 1.06$\times 10^{36}$ & DMC \\
J0901-4624 & 80.0 & 3032 & 4.00$\times 10^{34}$ & D \\
J0905-5127 & 221.0 & 1330 & 2.36$\times 10^{34}$ & D \\
B0906-49 & 112.0 & 1000 & 4.90$\times 10^{35}$ & DMEC \\
J1015-5719 & 38.6 & 2732 & 8.27$\times 10^{35}$ & DMEC \\
J1016-5857 & 21.0 & 3163 & 2.58$\times 10^{36}$ & DMEC \\
J1019-5749 & 128.0 & 10913 & 1.85$\times 10^{35}$ & D \\
J1020-6026 & 330.0 & 3277 & 9.59$\times 10^{34}$ & D \\
J1028-5819 & 90.0 & 1423 & 8.32$\times 10^{35}$ & DMEC \\
J1044-5737 & 40.3 & 1895 & 8.03$\times 10^{35}$ & DMEC \\
B1046-58 & 20.4 & 2900 & 2.00$\times 10^{36}$ & DMEC \\
J1052-5954 & 143.0 & 3143 & 1.34$\times 10^{35}$ & DM \\
J1055-6028 & 53.5 & 3830 & 1.18$\times 10^{36}$ & DMEC \\
J1101-6101 & 116.0 & 7000 & 1.36$\times 10^{36}$ & DMC \\
J1105-6107 & 63.2 & 2360 & 2.48$\times 10^{36}$ & DMEC \\
J1112-6103 & 32.7 & 4500 & 4.53$\times 10^{36}$ & DMEC \\
J1138-6207 & 149.0 & 7197 & 3.03$\times 10^{35}$ & D \\
J1151-6108 & 157.0 & 2216 & 3.87$\times 10^{35}$ & DMC \\
J1156-5707 & 172.0 & 2848 & 4.37$\times 10^{34}$ & D \\
B1259-63 & 332.0 & 2632 & 8.26$\times 10^{35}$ & DMEC \\
B1356-60 & 319.0 & 5000 & 1.21$\times 10^{35}$ & D \\
J1406-6121 & 61.7 & 7297 & 2.23$\times 10^{35}$ & D \\
J1410-6132 & 24.8 & 13510 & 1.01$\times 10^{37}$ & DMEC \\
J1412-6145 & 50.4 & 7115 & 1.25$\times 10^{35}$ & D \\
J1413-6205 & 62.8 & 2150 & 8.27$\times 10^{35}$ & DMEC \\
J1429-5911 & 60.2 & 1955 & 7.75$\times 10^{35}$ & DMEC \\
J1437-5959 & 114.0 & 8543 & 1.44$\times 10^{36}$ & DMEC \\
J1459-6053 & 64.7 & 1840 & 9.09$\times 10^{35}$ & DMEC \\
J1509-5850 & 154.0 & 3372 & 5.15$\times 10^{35}$ & DMEC \\
B1508-57 & 298.0 & 6835 & 1.27$\times 10^{35}$ & D \\
J1524-5625 & 31.8 & 3378 & 3.21$\times 10^{36}$ & DMEC \\
J1531-5610 & 96.7 & 2841 & 9.12$\times 10^{35}$ & DMEC \\
J1538-5551 & 517.0 & 5985 & 1.10$\times 10^{35}$ & D \\
J1541-5535 & 62.5 & 5161 & 1.14$\times 10^{35}$ & D \\
J1548-5607 & 252.0 & 5702 & 8.48$\times 10^{34}$ & D \\
J1549-4848 & 324.0 & 1308 & 2.32$\times 10^{34}$ & D \\
J1551-5310 & 36.9 & 5878 & 8.25$\times 10^{34}$ & D \\
J1601-5335 & 73.3 & 3576 & 1.03$\times 10^{35}$ & D \\
B1607-52 & 559.0 & 2949 & 3.36$\times 10^{34}$ & D \\
J1632-4757 & 240.0 & 4836 & 4.98$\times 10^{34}$ & D \\
J1636-4440 & 70.1 & 12455 & 2.09$\times 10^{35}$ & D \\
B1634-45 & 590.0 & 3442 & 7.52$\times 10^{34}$ & D \\
J1637-4642 & 41.2 & 4410 & 6.40$\times 10^{35}$ & DMEC \\
J1638-4608 & 85.6 & 4570 & 9.44$\times 10^{34}$ & D \\
J1643-4505 & 118.0 & 4738 & 9.39$\times 10^{34}$ & D \\
B1643-43 & 32.5 & 6226 & 3.58$\times 10^{35}$ & DM \\
J1648-4611 & 110.0 & 4468 & 2.09$\times 10^{35}$ & DM \\
J1702-4128 & 55.1 & 3971 & 3.42$\times 10^{35}$ & DM \\
J1705-3950 & 83.4 & 3426 & 7.37$\times 10^{34}$ & D \\
J1715-3903 & 118.0 & 3737 & 6.84$\times 10^{34}$ & D \\
J1718-3825 & 89.5 & 3488 & 1.25$\times 10^{36}$ & DMEC \\
B1718-35 & 176.0 & 4600 & 4.51$\times 10^{34}$ & D \\
B1719-37 & 344.0 & 2477 & 3.26$\times 10^{34}$ & D \\
J1723-3659 & 400.0 & 3497 & 3.80$\times 10^{34}$ & D \\
B1727-33 & 26.0 & 3488 & 1.23$\times 10^{36}$ & DMEC \\
J1737-3137 & 51.4 & 4162 & 5.99$\times 10^{34}$ & D \\
J1738-2955 & 85.7 & 3460 & 3.71$\times 10^{34}$ & D \\
J1739-3023 & 159.0 & 3074 & 3.01$\times 10^{35}$ & DMC \\
J1740+1000 & 114.0 & 1227 & 2.32$\times 10^{35}$ & DMEC \\
J1747-2958 & 25.5 & 2520 & 2.51$\times 10^{36}$ & DMEC \\
B1754-24 & 285.0 & 3124 & 4.00$\times 10^{34}$ & D \\
B1758-23 & 58.3 & 4000 & 6.20$\times 10^{34}$ & D \\
J1809-1917 & 51.4 & 3268 & 1.78$\times 10^{36}$ & DMEC \\
J1811-1925 & 23.3 & 5000 & 6.42$\times 10^{36}$ & DMEC \\
J1813-1246 & 43.4 & 2635 & 6.24$\times 10^{36}$ & DMEC \\
J1815-1738 & 40.4 & 4887 & 3.93$\times 10^{35}$ & DM \\
B1822-14 & 195.0 & 4440 & 4.11$\times 10^{34}$ & D \\
B1823-13 & 21.4 & 3606 & 2.84$\times 10^{36}$ & DMEC \\
J1828-1057 & 189.0 & 3648 & 5.47$\times 10^{34}$ & D \\
J1828-1101 & 77.2 & 4767 & 1.56$\times 10^{36}$ & DMEC \\
B1828-11 & 107.0 & 3147 & 3.56$\times 10^{34}$ & D \\
J1831-0952 & 128.0 & 3683 & 1.08$\times 10^{36}$ & DMEC \\
B1830-08 & 147.0 & 4500 & 5.84$\times 10^{35}$ & DMEC \\
B1832-06 & 120.0 & 5041 & 5.58$\times 10^{34}$ & D \\
J1835-0944 & 525.0 & 4215 & 5.64$\times 10^{34}$ & D \\
J1835-1106 & 128.0 & 3159 & 1.78$\times 10^{35}$ & DM \\
J1837-0604 & 33.8 & 4771 & 2.00$\times 10^{36}$ & DMEC \\
J1838-0453 & 51.9 & 6629 & 8.31$\times 10^{34}$ & D \\
J1838-0549 & 112.0 & 4061 & 1.01$\times 10^{35}$ & D \\
J1838-0655 & 22.7 & 6600 & 5.55$\times 10^{36}$ & DMEC \\
J1841-0345 & 55.9 & 3776 & 2.69$\times 10^{35}$ & DM \\
B1838-04 & 461.0 & 4399 & 3.91$\times 10^{34}$ & D \\
J1841-0524 & 30.2 & 4125 & 1.04$\times 10^{35}$ & D \\
J1846+0919 & 360.0 & 1530 & 3.41$\times 10^{34}$ & D \\
J1850-0026 & 67.5 & 6710 & 3.34$\times 10^{35}$ & D \\
J1853-0004 & 288.0 & 5339 & 2.11$\times 10^{35}$ & D \\
J1853+0056 & 204.0 & 3841 & 4.03$\times 10^{34}$ & D \\
B1853+01 & 20.3 & 3300 & 4.30$\times 10^{35}$ & DMEC \\
J1856+0245 & 20.6 & 6318 & 4.63$\times 10^{36}$ & DMEC \\
J1857+0143 & 71.0 & 4566 & 4.51$\times 10^{35}$ & DMC \\
J1906+0746 & 113.0 & 7400 & 2.68$\times 10^{35}$ & D \\
J1907+0918 & 38.0 & 8224 & 3.22$\times 10^{35}$ & D \\
J1909+0749 & 24.7 & 8286 & 4.50$\times 10^{35}$ & D \\
J1909+0912 & 98.7 & 7608 & 1.28$\times 10^{35}$ & D \\
J1913+0904 & 147.0 & 2997 & 1.60$\times 10^{35}$ & DM \\
J1913+1011 & 169.0 & 4613 & 2.87$\times 10^{36}$ & DMEC \\
J1916+1225 & 154.0 & 6486 & 7.87$\times 10^{34}$ & D \\
J1925+1720 & 115.0 & 5060 & 9.54$\times 10^{35}$ & DMEC \\
J1928+1746 & 82.6 & 4337 & 1.60$\times 10^{36}$ & DMEC \\
B1930+22 & 39.8 & 10900 & 7.54$\times 10^{35}$ & D \\
J1934+2352 & 21.6 & 12204 & 9.08$\times 10^{35}$ & D \\
J1935+2025 & 20.9 & 4598 & 4.66$\times 10^{36}$ & DMEC \\
J1938+2213 & 62.0 & 3419 & 3.66$\times 10^{35}$ & DMC \\
B1951+32 & 107.0 & 3000 & 3.74$\times 10^{36}$ & DMEC \\
J1954+2836 & 69.4 & 1960 & 1.05$\times 10^{36}$ & DMEC \\
J1958+2846 & 21.7 & 1950 & 3.42$\times 10^{35}$ & DMC \\
J2006+3102 & 104.0 & 6035 & 2.24$\times 10^{35}$ & D \\
J2021+4026 & 76.9 & 2150 & 1.16$\times 10^{35}$ & DM \\
J2032+4127 & 201.0 & 1330 & 1.52$\times 10^{35}$ & DM \\
J2238+5903 & 26.6 & 2830 & 8.89$\times 10^{35}$ & DMEC \\
J2240+5832 & 144.0 & 7275 & 2.21$\times 10^{35}$ & D \\
\hline
\caption{List of ATNF pulsars that would yield detectable Geminga-like haloes in the CTA GPS. The last column indicates the kind of constraints that could be obtained: D means plain detection only, M means true energy-dependent halo morphology can be separated from a 2D energy-independent Gaussian, E means extension can be detected beyond 30\,pc, and C means the true halo morphology can be separated from one with a 50\% higher suppressed diffusion coefficient.}
\label{tab:atnf-cand-geminga}
\end{longtable}
\clearpage
\twocolumn

\appendix

\section{Comparison between ctools and gammapy}
\label{app:ctools_gammapy}

In this appendix, we comment on the differences that occur between a gamma-ray analysis conducted with \texttt{ctools} and \texttt{gammapy} \cite{2017ICRC...35..766D}. Both software packages are publicly available and designed to perform scientific analyses on very high-energy gamma-ray data. While all results shown in the main text of this work have been derived with the functionalities of \texttt{ctools}, we repeated a small part of the analysis steps with templates generated by \texttt{gammapy} since it exhibits the same appeal as \texttt{ctools} and offers the opportunity to analyse data sets from other instruments than CTA. 

\begin{figure*}
\begin{centering}
\includegraphics[width=0.80\linewidth]{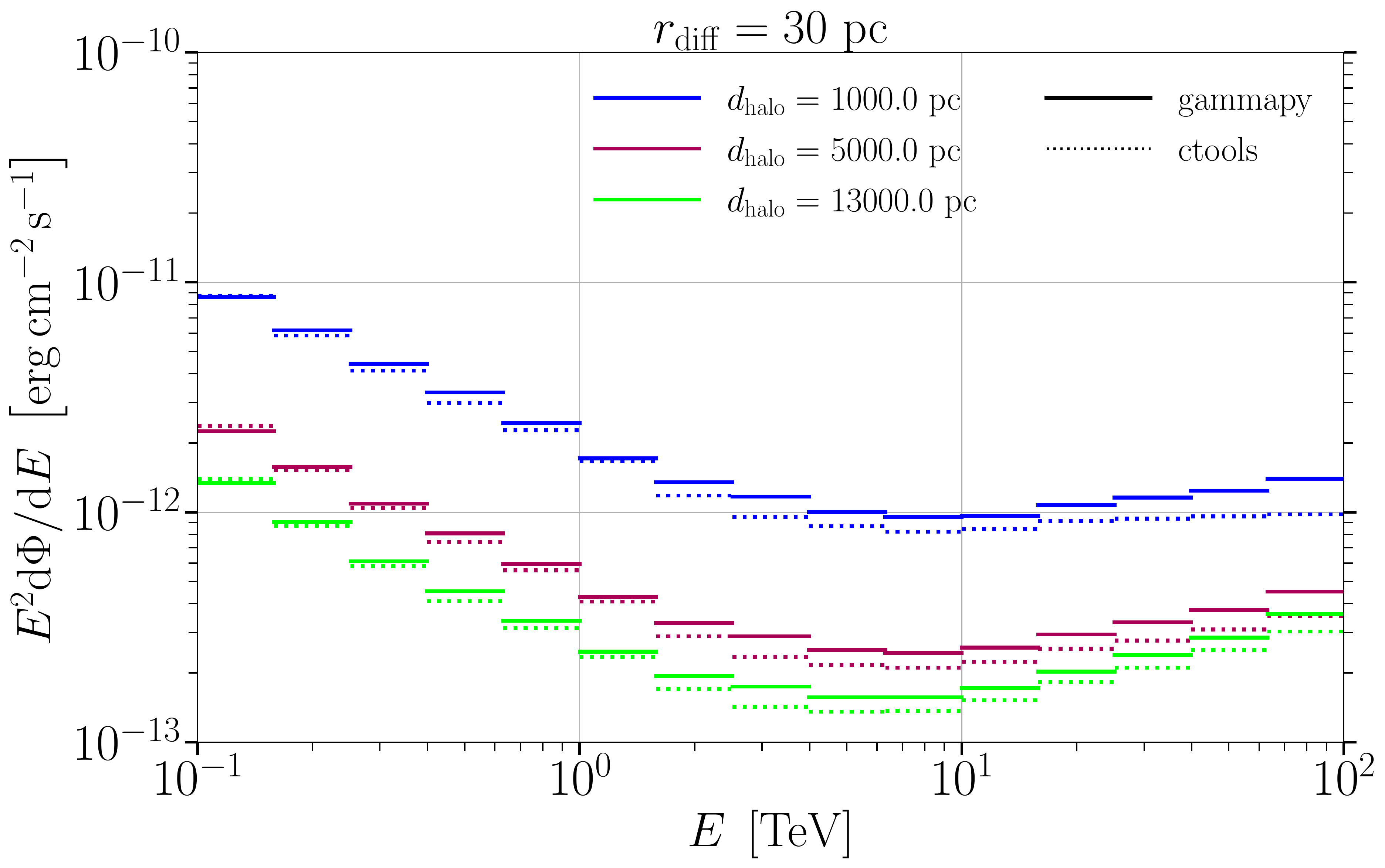}
\par\end{centering}
\caption{Differential spectral detection sensitivity ($5\sigma$) to Geminga-like pulsar haloes exhibiting a diffusion zone size of 30 pc located at $(\ell, b) = (-10^{\circ}, 0^{\circ})$ at various distances from the Earth studied under the conditions of CTA's Galactic plane survey. The solid sensitivity values represent the results obtained by using \texttt{gammapy} routines for the convolution with CTA's IRFs while the dotted values represent the equivalent scenario for templates generated with ctools.  \label{fig:ctools_vs_gammapy}}
\end{figure*}

To this end, we (re-)examine the $5\sigma$ detection sensitivity to a Geminga-like pulsar halo with a diffusion zone size of $r_{\mathrm{diff}} = 30$ pc located at various distances from the Solar system (cf.~Fig.~\ref{fig:DiffEnSens} (right)). The resulting sensitivity values are displayed in Fig.~\ref{fig:ctools_vs_gammapy}, where the solid lines represent the values obtained with \texttt{gammapy} while the dotted lines are the corresponding ctools equivalents. Up to energies of $\sim$1 TeV, both software packages agree reasonably well with each other. At higher energies, the results start to diverge in a way that the templates prepared with ctools lead to systematically better sensitivities. This effect may be as pronounced as $\sim20\%$ for the last energy bin and a halo at 1 kpc distance. Nonetheless, the profiles of the sensitivity curves are fairly similar. The existing differences might be explained with the explicit approach implemented in each software package to derive the Asimov prediction for a given flux model, which -- to the best of our knowledge -- is indeed different.

\section{The full GPS exposure maps}
\label{app:fullGPS}

\begin{figure*}
\begin{centering}
\includegraphics[width=0.8\linewidth]{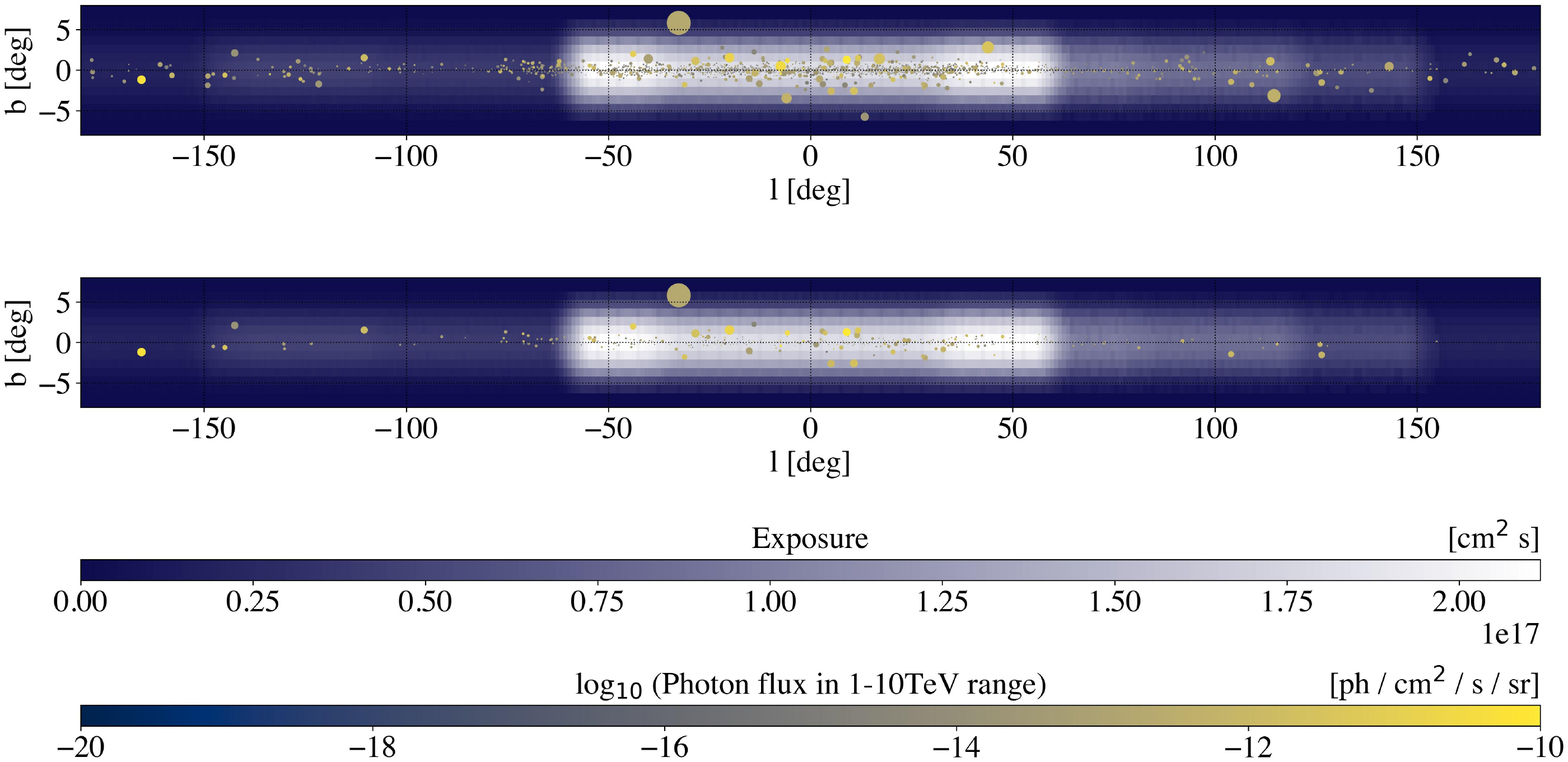}
\par\end{centering}
\caption{The exposure for the planned GPS observations overlaid with synthetic TeV halo population, showing the full Galactic plane. Note that, in order to preserve spherical shape of sources, they are plotted assuming the pixel size in the $b$ direction. The most sources from our synthetic halo population are in the region that has the deepest exposure. {\it Top:} All simulated sources. {\it Bottom:} Only resolved sources (see Sec. \ref{phy:popu}).} \label{fig:fullGPS}
\end{figure*}

In addition to Fig. \ref{fig:GPSexp} which shows the exposure of the planned GPS observations overlaid with synthetic TeV halo population over a limited longitude range, we show in Fig. \ref{fig:fullGPS} the exposure of the full GPS. The top (bottom) panels show all (only detected) TeV haloes within our synthetic population. The spherical shape of the TeV halo markers is preserved, with their size scaled in accordance with the latitude axis. The majority of the sources are in the central region which is expected to receive the most observation time and reach the deepest exposure.

%\vero{Should we comment on source confusion?}

\section{Impact from IE model}
\label{app:IEMfit}

\begin{figure*}
\begin{centering}
\includegraphics[width=0.48\linewidth]{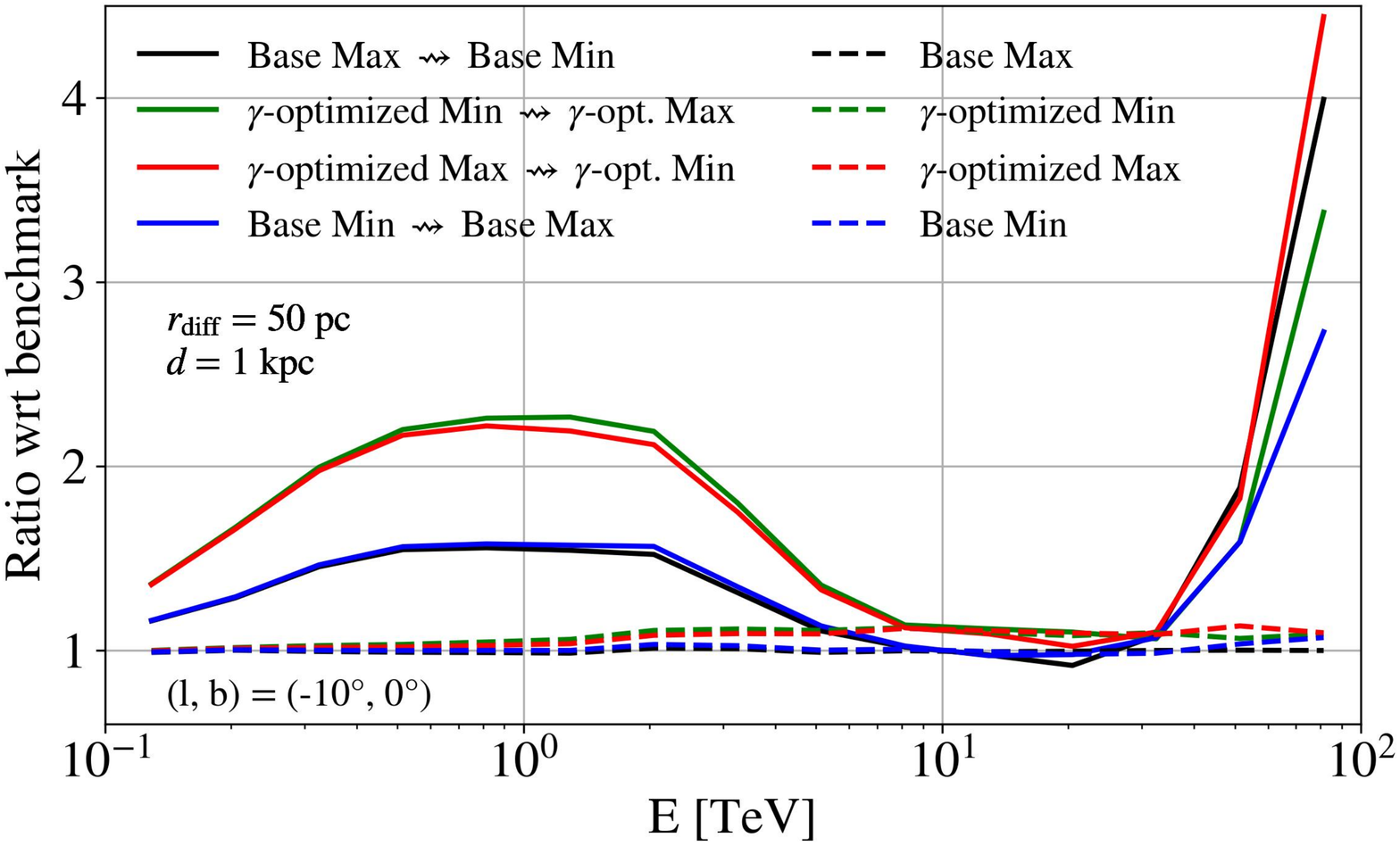}
\includegraphics[width=0.48\linewidth]{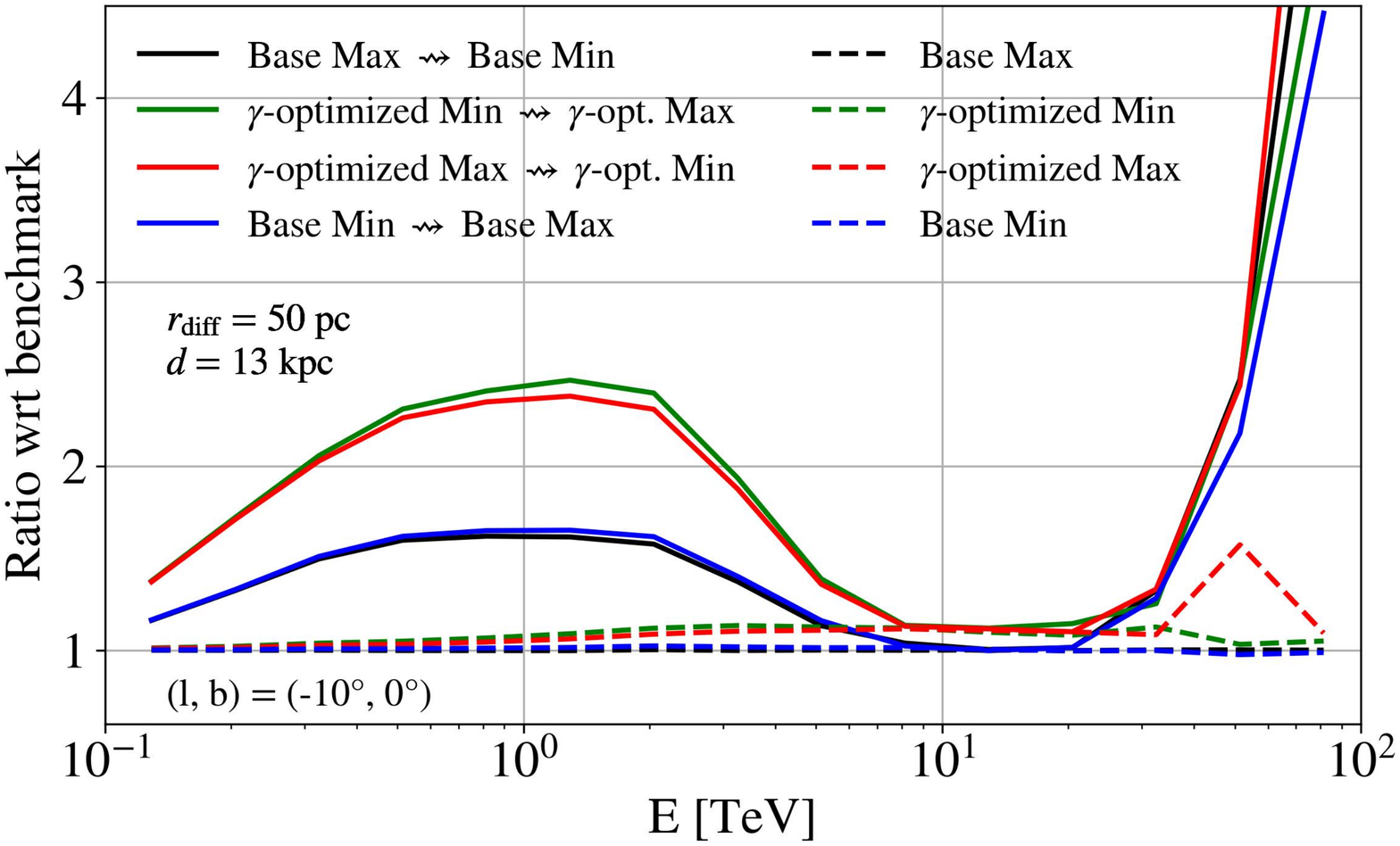}
\par\end{centering}
\caption{Impact of IE modeling on spectral sensitivity, with respect to the benchmark curve, for a pulsar halo model at a 1\,kpc (left) and 13\,kpc (right) distance. Dashed lines show the effect from using a different IE model throughout the whole process, and solid lines show the effect from using a model in sensitivity calculation that differ from the one used in data simulation. In the latter case, the model combinations correspond to the gray cells in Tab. \ref{tab:IEMfits}, and the first model in each line's label denotes the IE used in producing the mock data while the second represents the IE used in model fitting.}
\label{fig:IEMfits}
\end{figure*}

In a realistic analysis, the true IE properties are not known. In this appendix, we explore how using in model fitting a diffuse emission model other than the true one used in data simulation can affect the derived differential sensitivity. In a realistic data analysis, multiple IE models may be tested and fitted to the data, and the model yielding the best fit is selected for further analysis. We simulated mock data using different diffuse emission models and explored the log-likelihood variations from assuming different diffuse emission models. The resulting values are shown in Tab. \ref{tab:IEMfits}. The best-fit is naturally obtained when using in model fitting the same IE model that was used for producing the mock data, and we retain the corresponding results as a measure of the impact of the true IE properties. The second best-fit for each data set is highlighted in gray, and we retain the corresponding results as a measure of the impact of using an imperfect IE model on sensitivity. The comparison of the differential sensitivities with respect to our benchmark model (using Base Max IEM) for the best-fit and second best-fit IE combinations is shown in Fig. \ref{fig:IEMfits}. The impact of the different variants for IE modeling is very modest, while the worsening of sensitivity can reach up to a factor of 4 at the highest energies when using an inadequate model.

\clearpage
\onecolumn
\begin{table*}
    \begin{longtable}{| p{3cm} | p{2.5cm} | p{2.5cm} | p{2.5cm} | p{2.5cm} |}
    \hline
    \rowcolor{lightgray}\multicolumn{5}{|c|}{\textcolor{black}{Source at 1 kpc distance}} \\
    \hline
    Model | Data & Base Max & $\gamma$-optimized Min & $\gamma$-optimized Max & Base Min \\
    \hline
    Base Max & 0 & 502773 & 305625 & \cellcolor{lightgray}45001 \\
    $\gamma$-opt. Min & 483948 & 0 & \cellcolor{lightgray}95809 & 366297 \\
    $\gamma$-opt. Min & 299580 &\cellcolor{lightgray}97392 & 0 & 193519 \\
    Base Min & \cellcolor{lightgray}44422 & 376630 & 193474 & 0 \\
    \hline
    \multicolumn{5}{c}{} \\
    \hline
    \rowcolor{lightgray}\multicolumn{5}{|c|}{\textcolor{black}{Source at 13 kpc distance}} \\
    \hline
    Model | Data & Base Max & $\gamma$-optimized Min & $\gamma$-optimized Max & Base Min \\
    \hline
    Base Max & 0 & 64243 & 38223 & \cellcolor{lightgray}4763 \\
    $\gamma$-opt. Min & 62116 & 0 & \cellcolor{lightgray}10681 & 46699 \\
    $\gamma$-opt. Max & 38079 & \cellcolor{lightgray}10801 & 0 & 23917 \\
    Base Min & \cellcolor{lightgray}4824 & 47255 & 23157 & 0 \\
    \hline
    \end{longtable}
    \caption{Table of the log-likelihood value difference between a fit with the same IE as used to prepare the mock data and alternative IEMs, considering a source at 1 kpc (top) and 13 kpc (bottom) distance from the Sun. The second best-fit models for each mock data are highlighted in gray and the differential sensitivity with respect to the benchmark for these IE combinations is shown in Fig.~\ref{fig:IEMfits}.\label{tab:IEMfits}}
\end{table*}
\clearpage
\twocolumn

\end{document}